\documentstyle{mn}
\input psfig.tex
\newif\ifAMStwofonts

\ifoldfss
  \ifCUPmtlplainloaded \else
    \NewTextAlphabet{textbfit} {cmbxti10} {}
    \NewTextAlphabet{textbfss} {cmssbx10} {}
    \NewMathAlphabet{mathbfit} {cmbxti10} {} 
    \NewMathAlphabet{mathbfss} {cmssbx10} {} 
  \fi
  \ifAMStwofonts
    \ifCUPmtlplainloaded \else
      \NewSymbolFont{upmath} {eurm10}
      \NewSymbolFont{AMSa} {msam10}
      \NewMathSymbol{\upi}     {0}{upmath}{19}
      \NewMathSymbol{\umu}     {0}{upmath}{16}
      \NewMathSymbol{\upartial}{0}{upmath}{40}
      \NewMathSymbol{\leqslant}{3}{AMSa}{36}
      \NewMathSymbol{\geqslant}{3}{AMSa}{3E}

      \let\leq=\leqslant 
       
    \fi
  \fi
\fi 

\ifnfssone
  \newmathalphabet{\mathit}
  \addtoversion{normal}{\mathit}{cmr}{m}{it}
  \addtoversion{bold}{\mathit}{cmr}{bx}{it}
  \newmathalphabet{\mathbfit} 
  \addtoversion{normal}{\mathbfit}{cmr}{bx}{it}
  \addtoversion{bold}{\mathbfit}{cmr}{bx}{it}
  \newmathalphabet{\mathbfss} 
  \addtoversion{normal}{\mathbfss}{cmss}{bx}{n}
  \addtoversion{bold}{\mathbfss}{cmss}{bx}{n}
  \ifAMStwofonts
    \ifCUPmtlplainloaded \else
      \UseAMStwoboldmath
      \makeatletter
      \new@mathgroup\upmath@group
      \define@mathgroup\mv@normal\upmath@group{eur}{m}{n}
      \define@mathgroup\mv@bold\upmath@group{eur}{b}{n}
      \edef\UPM{\hexnumber\upmath@group}
      \new@mathgroup\amsa@group
      \define@mathgroup\mv@normal\amsa@group{msa}{m}{n}
      \define@mathgroup\mv@bold\amsa@group{msa}{m}{n}
      \edef\AMSa{\hexnumber\amsa@group}
      \makeatother
      \mathchardef\upi="0\UPM19
      \mathchardef\umu="0\UPM16
      \mathchardef\upartial="0\UPM40
      \mathchardef\leqslant="3\AMSa36
      \mathchardef\geqslant="3\AMSa3E

      \let\leq=\leqslant 

    \fi
  \fi
\fi 

\ifnfsstwo
  \DeclareMathAlphabet{\mathbfit}{OT1}{cmr}{bx}{it}
  \SetMathAlphabet\mathbfit{bold}{OT1}{cmr}{bx}{it}
  \DeclareMathAlphabet{\mathbfss}{OT1}{cmss}{bx}{n}
  \SetMathAlphabet\mathbfss{bold}{OT1}{cmss}{bx}{n}
  \ifAMStwofonts
    \ifCUPmtlplainloaded \else
      \DeclareSymbolFont{UPM}{U}{eur}{m}{n}
      \SetSymbolFont{UPM}{bold}{U}{eur}{b}{n}
      \DeclareSymbolFont{AMSa}{U}{msa}{m}{n}
      \DeclareMathSymbol{\upi}{0}{UPM}{"19}
      \DeclareMathSymbol{\umu}{0}{UPM}{"16}
      \DeclareMathSymbol{\upartial}{0}{UPM}{"40}
      \DeclareMathSymbol{\leqslant}{3}{AMSa}{"36}
      \DeclareMathSymbol{\geqslant}{3}{AMSa}{"3E}

      \let\leq=\leqslant 

    \fi
  \fi
\fi 

\ifCUPmtlplainloaded \else
  \ifAMStwofonts \else 
    \def\upi{\pi}
    \def\umu{\mu}
    \def\upartial{\partial}
  \fi
\fi

\def\simlt{\lower.5ex\hbox{$\; \buildrel < \over \sim \;$}}
\def\simgt{\lower.5ex\hbox{$\; \buildrel > \over \sim \;$}}
\def\ms{M$_{\odot}$}

\def\mp{M$_{\odot}$ pc$^{-2}$}
\def\mg{M$_{\odot}$ pc$^{-2}$ Gyr$^{-1}$}
\def\my{M$_{\odot}$ yr$^{-1}$}
\def\zs{Z$_{\odot}$}
\def\al{A$_{\lambda}$}

\title{Chemo-spectrophotometric evolution of spiral galaxies: \\
     I. The model and the Milky Way}
\author[S. Boissier and N. Prantzos]
       {S. Boissier and N. Prantzos \\
 Institut d'Astrophysique de Paris, 98bis, Bd. Arago, 75104 Paris}
\date{ }

\pagerange{\pageref{firstpage}--\pageref{lastpage}}
\pubyear{1998}

\begin{document}

\maketitle

\label{firstpage}

\begin{abstract}
The chemical and spectro-photometric evolution of spiral galaxies
is investigated
with detailed models, making use of up-to-date ingredients (like metallicity dependent
stellar properties) and a prescription for the star formation rate (SFR) justified both
empirically and theoretically. As a first application, the model is used to describe the
evolution of the Milky Way.
The role of the adopted scheme  of disk formation (``inside-out'') in shaping the
various chemical and colour profiles is investigated, as well as the role of extinction.
It is shown that the  solar neighborhood does not evolve as the Milky Way as a whole
and that one-zone models with a non-linear SFR prescription cannot be used to study the
evolution of our Galaxy. Our model average SFR is shown to match well observations
of external spirals.
\end{abstract}


\section {Introduction}

Galactic evolution is one of the major topics of research in modern astrophysics.
Despite more than three decades of intense observational and theoretical studies,
the subject has not yet reached the maturity of other fields, like e.g. stellar
evolution. Our limited understanding of the ``driver'' of galactic evolution
i.e. star formation from the interstellar medium, is the main reason for this
situation.

Galactic disks are privileged places for the study of the star formation rate (SFR)
since their various properties (surface densities of stars and gas, SFR,  abundances of
various elements, luminosity and colour profiles etc.) can be measured as a function of
galactocentric distance and compared to theoretical predictions. Notice that predictions
for the SFR exist only for disk galaxies (based on various instability criteria, e.g.
Talbot and Arnett 1975, Onihishi 1975, Wyse and Silk 1989, 
Dopita and Ryder 1994), not for ellipticals or irregulars. 
Testing these predictions  against
the largest observational data set is obviously crucial for our understanding of the SFR.

Until recently,  detailed observed  properties of disk galaxies were available
only for the local universe, i.e. at very low redshift. These data offer information
only about the current status of  galaxies, from which their past history is to be
derived; in view of the many model parameters, it is difficult to deduce a unique
history from those data (in one single case, namely the Milky Way, we have information
on the local disk history, via the age-metallicity relationship and the metallicity distribution
of low mass stars). The spectacular progress in instrumentation makes now possible
the observation  of galaxy morphology at higher redshifts (e.g. Lilly et al. 1998)
and will soon allow to compare  models to observations covering the largest part of galaxy
evolution.

The various aspects of galactic evolution, namely dynamical, chemical and 
spectro-photometric, have usually been treated separetely up to now (with a few exceptions).
A large number of multi-zone chemical evolution models have been proposed, matching
in a relatively satisfactory way the current chemical properties of the Milky Way disk
(e.g. Matteucci and Fran\c cois 1989, Ferrini et al. 1994, Prantzos and Aubert 1995,
Carigi 1996, Tosi 1998, Dwek 1998)
or the (much less constraining)  chemical profiles of external spirals (e.g. Molla et al. 1997);
in some cases, this class of models takes into account the possibility of radial flows
of gas in the galactic disks (e.g. Mayor and Vigroux 1981,
Lacey and Fall 1985, Clarke 1989, Tsujimoto et al. 1995,
Firmani et al. 1996).
On the other hand, one-zone spectrophotometric evolution models 
(treating the whole galaxy as one single region) have been developed,
matching the
integrated photometric properties of spirals along the Hubble sequence (usually varying the SFR
time-scale, e.g.  Guiderdoni and Rocca-Volmerange 1987, Bruzual and Charlot 1993, 
Arimoto et al. 1992);
these models do not provide a detailed study of the chemical properties or of the photometric
gradients in spirals.  Preliminary multi-zone
chemo-photometric evolution models have been studied by Talbott and Arnet (1975), while
more recent ones were applied to low surface brightness galaxies (e.g. Jimenez et al. 1998).

Chemo-dynamical evolution models have been developed and applied to spirals
by a few groups (Steinmetz and Muller 1994; Theis et al. 1997); these models have a
less sophisticated treatment of chemical evolution than the classical ones and
are still subject to the uncertainties of initial conditions and  gas-stars
interactions. In a recent work, Contardo et al. (1998) made the first attempt, to our
knowledge, to combine in a single computation photometry, chemistry and dynamics;
despite the failure to reproduce basic observational properties of the Milky Way disk
(like the local G-dwarf metallicity distribution - see Sec. 3.1.1 -
or colour gradients at low redshifts)
this type of model seems quite promising and shows the way that galactic evolution studies
should take in the future; however, in view of
the large and many uncertainties affecting dynamical
models (and their coupling to chemistry) at present, the way towards a convincing
unified treatment is expected to  be rather long.

This paper is the first of a series aiming to study in a unified framework the chemical
{\it and} spectrophotometric properties of spiral galaxies, with appropriate detailed
multi-zone  models. Here we present in some detail the model and its application to the Milky Way,
which will serve as ``template'' for further studies. In Sec. 2.1 we present the chemical
evolution part of the model, justifying our choice of the adopted SFR law (Sec. 2.1.2)
In Sec. 2.2 we present the spectro-photometric part of the model, emphasizing the homogeneity
of the adopted sets of stellar tracks and spectra (both as a function of metallicity)
and the importance of a correct treatment of metallicity-dependent stellar lifetimes in the
calculation of the photometric evolution.
In Sec. 3.1 we present the results of the model for the chemical and photometric evolution
of the solar neighborhood and study the effects of metallicity-dependent ingredients
and of extinction on those results.
In Sec. 3.2 we present the results for the Milky Way disk and we discuss the evolution
of the various chemical and photometric gradients and the (limited) role of extinction
in shaping the latter; we also discuss the populations of stellar remnants and their
role on various aspects of studies of the Galaxy. In Sec. 3.3 we compare the evolution of
the solar neighborhood to the one of the Milky Way as a whole and show that they differ
considerably, i.e. one-zone models reproducing the local disk properties cannot be used
to simulate the evolution of the whole Galaxy. In Sec. 3.4 we compare (favourably)
the average SFR
of the model with recent observations of the corresponding SFR in external spirals
(from Kennicut 1998). Finally, our main results are summarized in Sec. 4.

\section {The model}

\subsection {Chemical evolution}

The adopted model of galactic chemical evolution
has been presented in  previous works
(i.e. Prantzos and Aubert 1995, hereafter PA95; 
Prantzos and Silk 1998, hereafter PS98).
The classical set of the equations of galactic chemical evolution
(Tinsley 1980; Pagel 1997) 
is solved numerically for each zone, without the Instantaneous Recycling Approximation.
Instantaneous mixing is assumed within each zone, i.e. at the star's death
its ejecta  are thoroughly mixed in the local interstellar medium, which
is caracterized by a unique composition at any time (no abundance
scatter is obtained in that framework).
The main input  physics of the model are described below.

\subsubsection{Stellar inputs: yields, remnant masses, lifetimes}

The yields of massive stars (i.e. the amount of stellar ejecta in the form of
various elements) adopted in this work are those of Woosley
and Weaver (1995, hereafter WW95) for metallicity Z = \zs.
Notice that in some cases they differ considerably from the corresponding
ones of the other major recent work in the field 
(Thielemann et al. 1996). These
differences are due to the different physical inputs utilised by the
two groups, mainly the criterion for convection, the nuclear reaction
rates and  the
determination of the ``mass-cut''  (see Prantzos 1998b for a review of
the role of the yields of massive stars  in chemical evolution).
The most significant difference concerns the yields of  Fe:
in the case of WW95 they are increasing with stellar mass, while all
massive stars eject $\sim$0.07 \ms \  of Fe in the case of Thielemann et al.
(1996).

For the masses of the stellar remnants we have adopted the following
prescriptions. Stars with mass M $<$9 \ms \ become white dwarfs with
mass M$_R$(M/\ms) = 0.1 M/\ms+0.45 (Iben and Tutukov  1984).
Stars with M$>$8 \ms \ explode as accretion induced collapse 
(in the 8-11 \ms \ range) or
Fe core collapse (for M$>$11 \ms) supernovae, leaving
behind a neutron star of mass M$_R$ = 1.4 \ms (as suggested by observations
of neutron stars in binary systems, e.g. Thorsett and Chakrabarty 1998);
the heaviest of those
stars may form a black hole but the mass limit for its formation
is not known at present and cannot be inferred from theoretical or
observational arguments
(e.g. Prantzos 1994, Bethe and Brown 1995),
despite occasional claims for the contrary.
Fortunately, due to the steeply decreasing stellar Initial Mass Function
in the range of massive stars
(see Sec. 2.1.2), the mass limit for stellar black hole formation does not
significantly affect the results of chemical evolution, at least as far
as that limit  is above $\sim$40 \ms.
Attempts to determine the number and masses  of galactic neutron stars
and black holes
based on the energetics of the supernova explosion and the Fe core mass 
has been made in Timmes et al. (1996) and Bethe and Brown (1998); 
we feel however, that the current
understanding of these topics does not allow an accurate determination
of those parameters and we adopt the aforementionned simple prescription,
i.e.  stars with M$>$8 \ms leave a neutron star of 1.4 \ms, while
stars of M$>$40 \ms leave a black hole of 3 \ms \ (obviously, the adopted
values in the case of the black hole formation are  lower limits  and they
serve only for illustration purposes).

There are equally important uncertainties  in the yields of M $<$9 \ms
\ stars, due to the treatment of mass loss, enveloppe convection, etc.
especially in their final evolution on the Asymptotic Giant Branch (AGB;
see Charbonnel 1998 for a recent review). In this work we have adopted
those of Marigo et al. (1996) in the range 1-4 \ms \ and those of Renzini
and Voli (1981) in the 4-9 \ms \ range (their Table 4d with Hot
Bottom burning); despite the different physical inputs, the two sets of 
yields merge rather smoothly in the intermediate mass regime ($\sim$4 \ms).
In any case, the uncertainties in those yields play a very limited 
role in the context of this work.

Finally, the adopted stellar lifetimes $\tau_M$ as a function of the stellar mass $M$
are from the work of the Geneva group
(Schaller et al. 1992, Charbonnel et al. 1996).
dependence of the stellar lifetimes (see Fig. 3)
may affect considerably the results of photometric evolution (see Sec. 3.1.2).

\subsubsection{Stellar Initial Mass Function (IMF)}

The choice of the IMF plays a crucial role in the results of chemical
and photometric evolution of galaxies. Most photometric studies still use
a Salpeter power-law IMF $\Phi(M) \propto M^{-(1+X)}$ with $X$=1.35 in
the whole stellar mass range, although Salpeter (1955) derived that IMF in the
0.6-10 \ms \ range. However, several studies have shown that the IMF flattens
below M$\sim$1 \ms \ (e.g. Reid and Gizis 1997 and references therein)
and the use of a Salpeter IMF over the whole
mass range (i.e. down to the H-burning limit) is in disagreement with
observational evidence.

Although the  flatenning of the IMF below $\sim$1 \ms \ is
generally accepted now (e.g. Scalo 1998), its exact shape is still under debate.
For instance, Reid and Gizis (1997), using a new sample
of M-dwarfs and an up-to-date mass-luminosity calibration, find that
a single slope $X$=0.05 can fit the whole mass range 0.1 $<$ M/\ms$<$1.
In a recent review, Kroupa (1998), after discussing all recent determinations
of the low mass part of the IMF, favours a slope $X$=0.5  for
M$<$0.5 \ms \ and down to the H-burning limit.

On the other hand, the slope of the high mass part of the IMF is also
uncertain. Studies of OB associations on the SMC, LMC and the Milky Way
(Massey et al. 1995) show relatively flat slopes ($X\sim$1-1.5).
However, a surprisingly large number of massive stars ($\sim$50\%) are
found outside OB associations according to Massey (1998), who finds that
these ``field'' stars have much steeper IMFs ($X\sim$3-4 in the Milky Way,
LMC and SMC). The  slope of $X$=1.7  suggested in Scalo (1986) seems a reasonable
choice,    on the basis of all available observational data
 (see Gilmore et al. 1998 and references therein).

The mass limits of the IMF play also an important role in the
outcome of galactic evolution studies, in view of the normalisation
$$ \int_{M_{min}}^{M_{max}} \ \Phi(M) \ M \ dM \ = \ 1 \eqno(1) $$
$M_{max}$ is usually taken to be in the 50-100 \ms \ range
and its exact value affects slightly the results of chemical evolution.
The adopted $M_{min}$ is usually $\sim$0.1 \ms \ (i.e. close to the
H-burning limit of 0.08 \ms) but lower values cannot be excluded
provided they do not violate observational constraints.
In the case of the solar neighborhood, one such constraint is that
the curent surface density of brown dwarfs should not exceed $\sim$8 \mp
(Mera et al. 1998), i.e $\sim$15\% of the local surface density.
A very small amount of brown dwarfs is also suggested by Reid and Gizis (1997)
who find that their IMF drops below $\sim$0.1  \ms.

\begin{figure}
\psfig{file=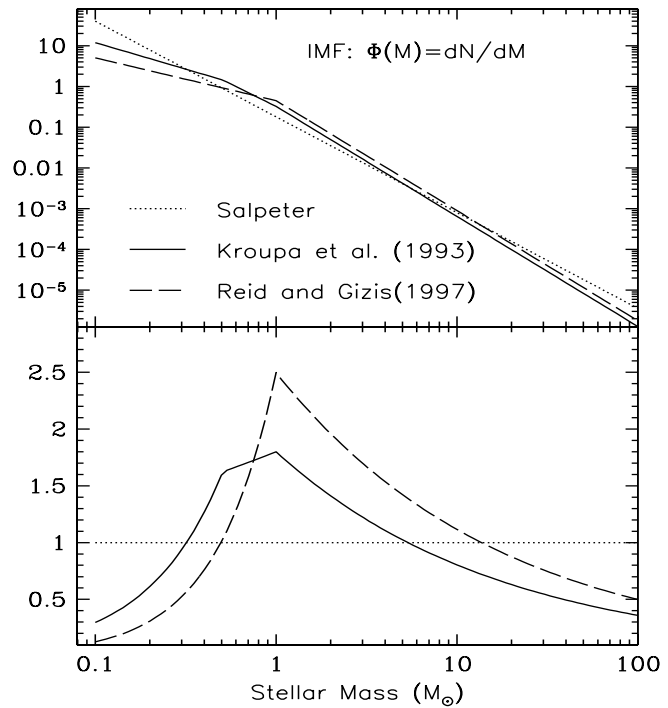,height=10cm,width=0.5\textwidth}
\caption{ \small
{\it Upper pannel:} Initial mass functions (IMF), 
according to Salpeter (1955, {\it dotted} curve with a unique slope $X$=1.35),
Reid and Gizis (1997, {\it dashed} curve) and Kroupa et al. (1993, {\it solid}
curve); the latter is adopted in this work.   All curves are normalised according
to Equ. 1 (see text).
{\it Lower pannel}: ratio of the multi-slope IMFs to the Salpeter IMF; because
of the normalisation, they contain more intermediate and low-mass stars
and  a smaller number of red dwarfs and massive stars.}
\end{figure}

These considerations suggest that there is some freedom in the choice
of the IMF. 
In our recent works (PA95, PS98) we
adopted the IMF  from the work of Kroupa et al. (1993, hereafter KTG93),
where the complex interdependence of several factors  (like stellar
binarity, ages and metallicities, as well as
mass-luminosity and colour-magnitude relationships) is explicitly
taken into account. It is  a three-slope power-law IMF ;
in the high mass regime it has
a relatively steep slope of $X$=1.7 (based on Scalo 1986),
 while it flattens in the low-mass
range ($X$=1.2 for 0.5$<$M/\ms$<$1. and  $X$=0.3 for M$<$0.5 \ms).
We adopt  here  again the KTG93 IMF
between 0.1 and 100 \ms, although we are aware that there is some debate as to the
exact form of the low-mass part (in view of the results of  Gould et al. 1997
and Reid and Gizis 1997; see, however, Haywood 1994).
The KTG93 IMF is plotted in Fig. 1, where it is
compared to the Salpeter IMF and the Reid and Gizis (1997) IMF. 
The KTG93 IMF has a smaller number of massive
stars than the Salpeter IMF and thereoff
it produces less metals and SNII (although  the return mass fraction is similar
in the two case: $\sim$0.30);
on the other hand, it contains a larger number
of intermediate and $\sim$solar mass  stars and thus produces more light
than the Salpeter IMF (since it is precisely the long-lived 1-2 \ms \ stars
that contribute most of the galactic light at late times). Notice that the
IMF suggested by Reid and Gizis (1997, with $X$=0.05 for M$<$1 \ms \ and $X$=1.7 for
M$>$1 \ms)  would produce even more light
(when appropriately normalised, as in Equ. 1).
These considerations have important implications for the
chemo-spectrophotometric evolution of galaxies, that are rarely considered;
in fact, most studies of photometric evolution adopt a Salpeter IMF in the whole
mass range, without worrying about local observational evidence or implications
for chemical evolution.

Finally, the question of the variation of the IMF in time or space has not received
a satisfactory answer for the moment, although most observers tend to favour the
``no variation'' option. For instance,
Massey et al. (1995) find no difference in the IMF  of massive stars in
OB associations between the SMC, LMC and the Milky Way, despite the factor of $\sim$10
difference in metallicity between those regions; also, in a recent work
Gizis and Reid (1999) find that the low-mass part of the IMF is the same in local
disk and halo stars, i.e. independent of metallicity. On the other hand, Scalo (1998)
rejects the ``myth of the universality of the IMF'', on the basis of observational and
theoretical arguments
(for a thorough discussion of all aspects of the IMF see the volume edited by
Gilmore et al. 1998).

\begin{figure}
\psfig{file=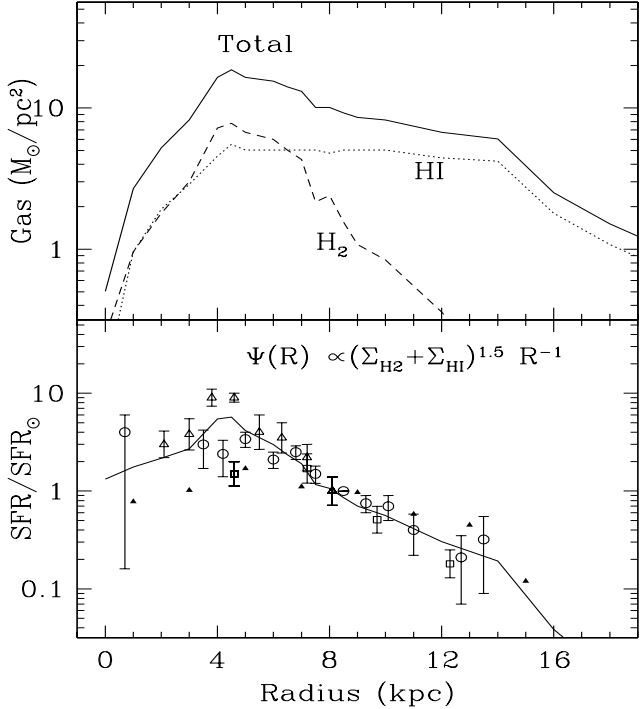,height=10cm,width=0.5\textwidth}
\caption{ \small
{\it Upper pannel:} Current surface density profiles of molecular (H$_2$) and
atomic (HI) hydrogen in the Milky Way, as a function of galactocentric radius
(from Dame 1993) and total gas surface density (the sum of the two, increased
by 40\% to account for helium).
{\it Lower pannel}: Corresponding theoretical current SFR ({\it solid} curve,
 according to Equ. 2) and comparison to observational estimates of the
current SFR profile. Data from: Lyne et al. 1985 (open circles), 
Gusten and Mezger 1983 (triangles),  Guibert et al. 1978 (squares).
The SFR profiles are normalised to their value at $R_S$=8 kpc.}
\end{figure}

\subsubsection{Star formation rate (SFR)}

The adopted star formation rate (SFR) $\Psi$ is locally a Schmidt type
law $\Psi \propto \Sigma_G^{1.5}$, where $\Sigma_G$ is the local
gas surface density. Such a proportionality has been suggested by Kennicut (1989, 1998)
on the basis of observations of average SFR vs. gas surface densities in spirals
(see Sec. 3.4). On the other hand, a radial dependence of
the star formation efficiency in disk galaxies is required in order
to reproduce the observed gradients, and such a dependence
has indeed been proposed   on the basis of various
instability criteria for gaseous disks  (e.g. Talbot and Arnett 1975;
Onihishi 1975; Wyse and Silk 1989; Dopita and Ryder 1994). It should be noted that
star formation theories exist and may be tested mainly for disk
galaxies, not for e.g. ellipticals or irregulars. 
We adopt in this work a star formation rate  explicitly dependent on
galactocentric radius. It is
based on the idea that stars are formed
in spiral galaxies when the interstellar medium with angular frequency
$\Omega(R)$ is periodically compressed by the passage of the spiral
pattern, having a frequency $\Omega_P$=const.$<< \Omega(R)$. This leads
to SFR $\propto \Omega(R)-\Omega_P \propto \Omega(R)$ and,
for disks with flat rotation curves, to SFR $\propto$ R$^{-1}$
(Wyse and Silk 1989), i.e.
$$ \Psi(t,R) \ = \ 0.1 \ \Sigma_G(R)^{1.5} \ (R/R_S)^{-1} 
\ M_{\odot} pc^{-2} Gyr^{-1} \eqno(2) $$
where $R_S$=8. kpc is the distance of the Sun to the Galactic centre.
Such a radial dependence of the SFR is also compatible with observational
evidence, as can be seen in Fig. 2, displaying
the current gas surface density profile in the Milky Way disk
(upper panel) and the corresponding SFR
according to Equ. 2 (lower pannel); comparison of this theoretical SFR to observations
(lower panel in Fig. 2) shows a fairly good agreement. 
To our knowledge, this is the first time that this type of direct
comparison is pointed out. It has been shown that this form of the SFR can
account for the observed gradients of gas fraction, SFR and chemical abundance
profiles in the Milky Way (e.g. PS98 and Sec. 3.2.1).
One of the major aims of the present work is  to derive the corresponding
photometric gradients (Sec. 3.2.2) and compare our model average SFR
to recent observations concerning external spirals (Sec. 3.4).

\subsubsection{Supernova rates}

Once the SFR and IMF are fixed, and the lower mass limit $M_{SNII}$ for
the formation of Fe core collapse supernova is determined, the
corresponding rate of SNII explosions can be calculated as
$SNII(t) \ = \ \Psi(t) \ \int^{100 M_{\odot}}_{M_{SNII}}  \Phi(M)  dM$; we
adopt here M$_{SNII}$=8 \ms (in fact, the most massive of these stars,
as well as some of those in close binary systems, will suffer extensive
mass losses prior to the explosion and appear as SNIb/c supernovae; but
we keep the designation of SNII here for all  core collapse events).
With the adopted IMF, the integral in the formula above has the value
5.5 10$^{-3}$, i.e. 0.55 SNII per century are expected for a
SFR of 1 \ms \ year$^{-1}$; in the case of a Salpeter IMF, the corresponding
number is slightly higher (7.7 10$^{-3}$).

On the other hand,
a supplementary source of Fe peak nuclei is introduced in the form of type
Ia supernovae (SNIa), presumably  white dwarfs in binary systems
accreting material from their companion star (see Nomoto et al. 1997
for a recent review of SNIa).
The relative homogeneity of the  observed   SNIa
lightcurves suggests that            $\sim$0.7 \ms \ of Fe are
ejected (originally in the form of $^{56}$Ni); this is also predicted
by the most successful models of SNIa, involving carbon deflagration
in a Chandrasekhar mass white dwarf. We adopt here the yields of the
W7 model from Thielemann et al. (1986).
However, the evolution of the  SNIa  rate in
the Galaxy is difficult to predict from the theory of binary systems
alone (see e.g. Ruiz-Lapuente et al. 1997 and references therein).
We adopt here the prescription of Matteucci and Greggio (1986), in order to
reproduce the observed decline of O/Fe vs Fe/H in the local disk. Notice that the recent
results of Israelian et al. (1998) and Boesgaard et al. (1999) show a steady
decline of that ratio also in halo stars, in disagreement with previous observations
showing a "plateau"; if these new results are confirmed, the SNIa rate formalism
should be reconsidered.

The comparison of the model supernova rate (integrated over the disk)
to observations of Milky Way type spirals (usually
expressed in SNU, i.e. number of supernovae per century and per 10$^{10}$
L$_{B,\odot}$) is a crucial test of the model. It is feasible only
in the framework of self-consistent models calculating both chemical
{\it and} photometric evolution and has rarely been done up to now (see
Sec. 3.3).

\subsubsection{Gaseous flows}

The disk is assumed to be
built by accretion of gas with primordial composition. On purely
phenomenological grounds, infall constitutes the most elegant and natural
way to account for the G-dwarf problem in the solar neighborhood 
(e.g. Tinsley 1980,
Pagel 1997), but is also supported by some chemodynamical models
of the Milky Way (Samland et al. 1997). 

The infall rate $f(t,R)$ is normalised to  the local 
disk surface density $\Sigma_T(R)$: 
$$\int_0^T \ f(t,R) \ dt \ = \ \Sigma_T(R) \eqno(3) $$
where $T$=13.5 Gyr is the adopted age of the disk.
The form of $f(t,R_S=8 kpc)$ is adjusted to satisfy  the constraint of the
G-dwarf metallicity distribution in the solar neighborhood.
An exponentially decreasing
infall rate $f(t)\propto e^{-t/\tau}$
with $\tau>$7 Gyr can provide a satisfactory fit to the new
data (Chiapini et al. 1997 and 
Sec. 3.1.1), and we adopt here this minimal value for $\tau(R_S$).
We note that such long time scales for the disc formation (many Gyr)
are also obtained in recent chemodynamical models  (Samland et al. 1997).
To simulate the
``inside-out'' formation of the disk (suggested by dynamical models,
e.g. Larson 1976), the  infall timescale $\tau(R)$
is assumed to be radially dependent, taking lower values
in the inner regions ($\tau$=1 Gyr at radius R=2 kpc)
and larger ones in the outer disk ($\tau$=10 Gyr  at R=18 kpc).
 The radial variation in the SFR efficiency and in the infall 
timescale are
 the   only parameters of the model explicitly dependent on radius.

We assume that the disk is evolving as independent one-zone rings.
This (over)simplification
ignores in general the possibility of radial inflows in gaseous disks, 
resulting e.g. by viscosity  or by
infalling gas with specific angular momentum different from the one
of the underlying disk; in both cases, the resulting redistribution of angular
momentum leads to radial mass flows. The magnitude of the effect is
difficult to evaluate, because of our poor understanding of viscosity 
and our ignorance of the kinematics of the infalling gas.
Models with radial inflows have been explored in the past
(Mayor and Vigroux 1981; Lacey and Fall 1986; Clarke 1989; 
Chamcham and Tayler 1994); at the present
stage of our knowledge, introduction of radial inflows in the models
 would imply even more free parameters and make
impossible  the study of a radial variation in the efficiency of the SFR. 

\subsection {Spectrophotometric evolution}

Once the chemical evolution of a galactic zone has been calculated, its
spectro-photometric evolution can be also follwed. The spectrum
of the zone at time $t$ is the sum of the individual spectra 
$l_{\lambda}(M,t-t',Z(t'))$ of stars of mass $M$, formed at time $t'<t$ with
metallicity $Z(t')$ and still alive at time $t$:
$$ {\rm L}_{\lambda}(t)   = \int_t  \int_M  \Psi(t')
  \Phi(M)   l_{\lambda}(M,t-t',Z(t'))   dM   dt' \eqno(4) $$
This integral can, in principle, be evaluated by integrating
first either on time (isomass method) or on mass (isochrone method).
The limits of the corresponding integrals are not the same in the two
methods. In the first one:
$$ {\rm L}_{\lambda}(t)   = \int_{M_{min}}^{M_{max}} \Phi(M) [ \int^t_{t_{inf}(Z(t'))}
\Psi(t')  l_{\lambda}(M,t-t',Z(t')) dt' ] dM \eqno(4a) $$
\noindent while in the second one:
$$ {\rm L}_{\lambda}(t)   = \int_0^t \Psi(t') [ \int_{M_{min}}^{M_{up}(Z(t'))} \Phi(M)
 l_{\lambda}(M,t-t',Z(t')) dM ] dt' \eqno(4b) $$

In the first case (isomass), the lower limit of the first integral (inside
[ ]) is $t_{inf}(Z(t'))=t-\tau_M(Z(t'))$ and represents the creation time
of the star of mass $M$ dying at time $t$. This limit
depends itself on the integration variable $t'$ of the first integral
because of the metallicity
dependent lifetimes of stars; obviously, in that case the first integral
cannot be calculated and the isomass method cannot
be applied. In the second case (isochrone), the upper limit of the first
integral (inside [ ]) is $M_{up}(Z(t'))=M[\tau_M(Z(t'))=t-t']$, i.e.
it corresponds to the heaviest star created at time
$t'$ and dying  at time $t$; this limit does not depend on the integration
variable $M$ and the integration can always be performed. Thus,
although the two methods are, in principle, equivalent when
metallicity independent lifetimes are considered,
{\it the isochrone method is the only applicable when
metallicity dependent lifetimes are taken into account} (as they should).
We adopt the second method in this work.

Calculation of the double integral in eqs. (4) requires knowledge of:

a) the stellar IMF $\Phi(M)$;
b) the SFR $\Psi(t)$; 
c) the metallicity $Z(t)$; 
d) the stellar spectra $l_{\lambda}(M,t,Z)$ for all evolutionary
stages and initial metallicities.

Ingredient (a) is the same as the one adopted in Sec. 2.1.2, ingredients
(b) and (c) are obtained self-consistently, by the adopted prescription
for the SFR and the run of the chemical
evolution model and ingredient (d) is discussed in Sec. 2.2.1 and 2.2.2
below.

Eq.  (4b) gives the composite absorption spectrum of the stellar
population of the galaxy. This spectrum is subsequently modified by
ionisation of the ambient gas (its UV part) and absorption by dust.
The (crude) modelisation of the latter process is discussed in  Sec.
2.2.3.

\begin{figure*}
\psfig{file=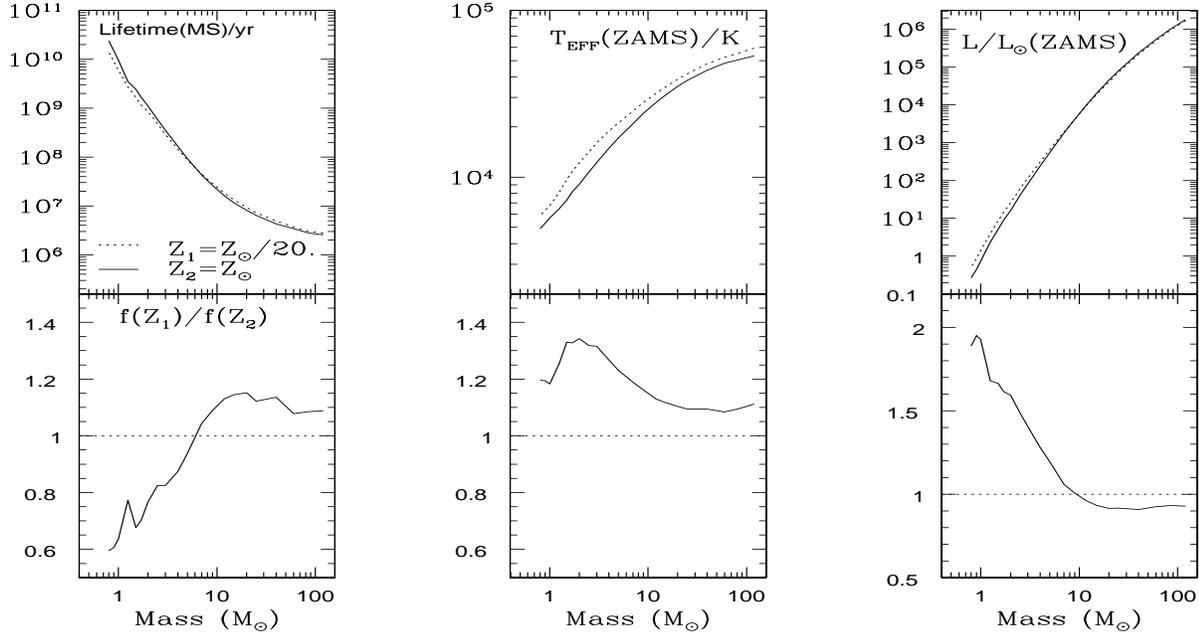,angle=-90,height=9cm,width=\textwidth}
\caption{ \small
Comparison of Main Sequence (MS) lifetimes ({\it left }), Zero-Age Main Sequence
(ZAMS) effective temperatures ({\it middle}) and luminosities ({\it right})
of stars with metallicity $Z_1$=\zs/20  ({\it dotted} curves in the {\it upper}
pannels) and $Z_2$=\zs \ ({\it solid} curves in the {\it upper } pannels).
In the {\it lower} pannels is displayed the ratio of the corresponding
quantities f($Z_1$)/f($Z_2$). Stars with lower metallicities are in general hotter;
those of mass $<$5\ms \ are more luminous and have shorter lifetimes
 than their higher
metallicity counterparts. All data are from the library of the Geneva stellar
evolution tracks (see Sec. 2.2.1).
}
\end{figure*}

\subsubsection{ Stellar evolution tracks }

We adopt in this work the tracks of the Geneva group (Schaller
et al. 1992; Charbonnel et al. 1996), concerning stars in the mass
range 0.8 $\leq$ M/\ms $\leq$ 120 and in the metallicity range
$Z_1$=0.05 \zs \ to $Z_2$=\zs.
The positions of the stars in the HR diagram
(effective temperature $T_{eff}$ and bolometric luminosity $L$) are given,
along with the corresponding ages, at equivalent points of the
evolution (e.g. main-sequence turn-off, He ignition, etc.), allowing
for an easy interpolation to other stellar masses and metallicities
in the given range. 

Stars with M$<$0.8 \ms \ and main sequence
lifetimes $\tau_{MS} >$13 Gyr are always attributed their Zero-Age
Main Sequence  (ZAMS) position, assuming an appropriate mass-luminosity
relationship (Baraffe et al. 1997);  in view of their very low ZAMS luminosity, their
role is negligible in the framewrok of this work.
For $Z < Z_1$  we adopt the tracks for $Z=Z_1$ and for
$Z > Z_2$ the tracks for $Z_2$. This approximation has a negligible
effect on the final results, since very few stars in the disk have
metallicities outside that range (see Sec. 3.1 and 3.2).

The metallicity dependence of the stellar tracks and
lifetimes may have a considerable impact on the photometric evolution.
For that reason, we present in Fig. 3 the main sequence lifetimes,
and the zero-age main sequence T$_{eff}$ and $L$ of the adopted
models for Z=\zs \ and Z=\zs/20. Stars of low-mass and low metallicity
may be $\sim$30\% hotter, $\sim$60\% more luminous and have $\sim$30\%
shorter lifetimes than their more ``metallic'' counterparts.

\subsubsection {Stellar spectra}

An important contribution to studies of galactic chemical evolution
has been recently made by Lejeune et al. (1997). They constructed a
homogeneous library of synthetic stellar spectra by compiling
different sets of stellar atmosphere  models published in the litterature.
The library covers the
wavelength range from 9 nm to 160 $\mu$m with a resolution of 
1 nm in the UV, 2 nm in the optical and 10 nm in the infrared.
Stellar spectra are given in the 3-dimensional phase-space of
$T_{eff}$, surface gravity g and metallicity [Fe/H], and are
uniformly sampled in the range 2500 K $< T_{eff} <$ 50000 K,
-1 $<$ log(g) $<$5.5 and -3.5$<$[Fe/H]$<$+1. 
In the temperature range 3 500 K to 50 000 K the models are identical to
those of Kurucz (1995). In the range 2 500 K to 3 500 K, composite
models of M giants with [Fe/H] between -1.5 and +0.5 were constructed,
by connecting the Fluks et al. (1994) synthetic spectra with the Bessel et
al. (1991) models. Finally, M-dwarf model spectra in the range
2 000 K to 3 500 K were selected from the grid of Allard and 
Hauschildt (1995). As described in Lejeune et al. (1997), these
synthetic spectra have been corrected as to
reproduce empirical colour-magnitude relationships, thus eliminating
the risk of theoretical errors.

The homogeneity of the
library greatly facilitates the calculation of stellar luminosities
$l_{\lambda}$, allowing an easy interpolation in
stellar mass, metallicity and evolutionary stage.
Magnitudes and colours are calculated  in the standard
photometric system UBVRIJK  for each galactic zone.

\subsubsection{Extinction by dust}

Interstellar dust absorbs
light from the stars  with a wavelength dependent efficiency
and re-emits it in the far infrared (FIR) range.
The resulting extinction as a function of wavelength is
difficult to evaluate from
first principles (see Corradi et al. 1996 for a recent attempt, solving
the full radiative transport problem in the case of  a disk) and
models usually adopt empirical prescriptions.

Extinction, i.e. the quantity \al=-2.5 log(L$_{\lambda}$/L$^0_{\lambda}$)
where L$^0_{\lambda}$ is the intrinsic luminosity of the stellar
population calculated in eq. (3) and L$_{\lambda}$ the emerging one,
depends on both the
physical processes involved (absorption and scattering by dust grains)
and the geometry of the source.
The former is usually parametrized with the introduction
of an effective optical  depth $\tau_{\lambda}$
depending on the amount of dust and the absorbing and reflecting
properties of grains, while the latter depends on the assumed
relative distributions of stars and dust.

The effective optical depth is expressed in terms of the exctinction
properties observed in the solar neighborhood: i) the extinction in the
V band per gas surface density $(A_V/N_H)_{\odot}$=5.3 10$^{-22}$ mag cm$^2$
(Bohlin et al. 1978)
where $N_H$ is the surface density of H (in atoms per cm$^2$); and ii)
the mean extinction curve $(A_{\lambda}/A_V)_{\odot}$, adopted here from
Natta and Panagia (1984).
The dependence on metallicity is introduced by a wavelength independent
factor $(Z$/\zs)$^s$, with $s$=1.35 for $\lambda<$2000 \AA \ and $s$=1.6
for $\lambda>$2000 \AA \ (adopted from Guiderdoni et al. 1998).
The effective optical depth is expressed as:
$$
\tau_{\lambda}= 0.92\  a_{\lambda} \left( \frac{A_\lambda}{A_V}\right)_{\odot}\left
(\frac{A_V}{N_H}\right)_{\odot} \left(\frac{Z_g}{Z_{\odot}}\right)^s N_H  \eqno(5)
$$

\noindent where a$_{\lambda}$ represents a positive contribution to the emerging radiation
from light scattered by dust grains, having an albedo $\omega_{\lambda}$.
We adopt in this work the approximation of Calzetti et al. (1994), concerning
a case intermediate between isotropic scattering and purely forward scattering;
it turns out that $\alpha_{\lambda}$ is given then by:
$$
a_{\lambda}=h_{\lambda} \sqrt{1-\omega_{\lambda}}+(1-h_{\lambda})(1-\omega_{\lambda}) \eqno(6)
$$
where $h_{\lambda}$ is a phase factor given by:
$$
h_{\lambda}=1 - 0.561 \ exp\left(-\frac{|log(\lambda)-3.3112|^{2.2}}{0.17}\right) \eqno(7)
$$
with $\lambda$ expressed in \AA. The grain albedo $\omega_{\lambda}$ in Calzetti et
al. (1994) and in this work is
 adopted from the work of Natta \& Panagia (1984).
(For further discussion on the derivation of the above formulae, the reader is
refered to Calzetti et al. 1994).

The geometry of the distribution of dust is taken into account as in 
Xu et al (1997), in a ``sandwich model'': 
one half of the stars is assumed to be mixed
with the layer of dust, while the other half lies on each side of this layer,
the quarter behind the dust being obscured by the screen. The
resulting extinction is then:
$$
A_{\lambda}=-2.5 \ log \left(0.25 + 0.5 \frac{1-e^{-\mu}}{\mu} +
0.25 e^{-\mu} \right) \eqno(8)
$$
where $\mu=\tau_{\lambda}/cos(i)$ and $i$ is the inclination angle
($i$=0$^o$ for face-on galaxies). The first term in parenthesis represents
contribution to the emerging luminosity from the stars in front of
the dust layer, the second term from the stars mixed with dust and the third one
from the stars behind the dust layer.

Extinction should, in principle, be treated with a full radiative
transfert code, as e.g. in Corradi et al. (1996). They find that the effect of
forward scattering is a very important one, leading to small values of extinction
even for large values of the optical depth (e.g. $A_V\sim$0.1 mag for
$\tau_V\sim$4). We are fully aware that
the prescription adopted here is a very crude one (although it is being
routinely adopted in many studies of that kind). For that reason we 
present below systematically our results for both cases:
without (i.e. stellar population only) and with extinction
taken into account, according to the adopted prescriptions. In a forthcoming
paper we explore the consequences of the radiative transfer treatment of
Corradi et al. (1996) on the absorption in various wavelengths.

\def\S{$\Sigma$}
\def\mp{M$_{\odot}$ pc$^{-2}$}
\def\mg{M$_{\odot}$ pc$^{-2}$ Gyr$^{-1}$}
\def\lb{L$_{B\odot}$}
\def\lv{L$_{V\odot}$}
\def\lk{L$_{K\odot}$}
\def\p2{pc$^{-2}$}

\begin{table*}


{\bf Table 1:} Main observational data for the solar neighborhood and results of a simple model
\begin{tabular}{ l l c c l }
\hline
Observable & Observed      & Reference  & Computed  & Main relevant parameter \\
           &  value        &            &  value    &    in the model         \\
\hline
                    &                         &      &  &   \\
{\underline {Surface densities of:}} &        &      & &  \\
 Gas                & \S$_G$=13$\pm$3 \mp  & 1& 12.0 &  Star Formation History \\
 Stars (visible)    & \S$_*$=35$\pm$5 \mp  & 2& 34.5 &                      \\
 Stars (visible+remnants)& \S$_*$=43$\pm$5 \mp  & 3& 39.9 & [ Stellar IMF \\
 Total (stars+gas)  & \S$_T$=51$\pm$6 \mp  & 4& 53.0 &                \\
                    &                      &  &  &          +    \\
 Gas fraction       & $\sigma_G$=0.15-0.25 &  & 0.22 &  Infall Rate   \\
                    &                      &  &   &   \\
 Star formation rate& $\Psi_o$=2-5 \mg     & 5& 3.2 &       +      \\
 Past average SFR   & $<\Psi> \sim$3 \mg   &   & 2.6 &  Star Formation Law ]  \\
 SNII rate          & 0.02 pc$^{-2}$ Gyr$^{-1}$ & 6 & 0.018 & \\
                    &                      &  & &   \\
 Present Day Mass Function & Low mass part: uncertain&7,8  & & IMF + SFR  \\
 (PDMF)             & (flatter than $X$=1.35)      &  & &   \\
                    &                      &  & &   \\
 {\underline {Abundances}} &               &  & &   Stellar yields \\
 At T$_0$-4.5 Gyr   & X$_{i,\odot}$        &9,10  & &                  \\
                    & X$_{O,\odot}$=9.20 10$^{-3}$  &  &7.6 10$^{-3}$ &    \\
                    & X$_{Fe,\odot}$=1.17 10$^{-3}$  &  &1.1 10$^{-3}$ &    \\
 At T$_0\sim$13.5 Gyr & X$_{i,0} \sim$ X$_{i,\odot}$      & 11, 12  &              &                  \\
                    &                      & &                &   \\
 O/Fe  vs  Fe/H :  & Decline for [Fe/H]$>-1$&13  & &   SNIa rate      \\
                    &                      &  & &   \\
 Age-metallicity  & Slow increase of Z     &13  & &   Star formation + yields \\
   (Z vs. t)        &  Dispersion (?)      &  & &   \\
                    &                      &  & &   \\
 Metallicity distribution &  Narrow        &14,15  & &   Infall rate \\
 of G-dwarf stars        & Peaked at [Fe/H]$\sim$-0.1     &  & &               \\
                    &                      &  & &   \\
{\underline {Luminosities:}} &L$_B$=20.0$\pm$2.0 \lb \p2  &16  &26.8 &  Star Formation History \\
                             &L$_V$=22.5$\pm$3.0 \lv \p2  &17  &25.9 &          +        \\
                             &L$_K$=68.0$\pm$23. \lk \p2  &18  &85.36 & Stellar tracks \\
                    &                      &  & &     Stellar Spectra \\
{\underline {Colours:}} &B-V = 0.85$\pm$0.15 &16  &0.71 &  Extinction   \\
                    & \ \ \ \ \ \ \ \ \ (0.63$\pm$0.1)     & 19 & &   \\
                    &                      &  & &   \\
\hline
\end{tabular}

References:
1. Kulkarni and Heiles 1987 ;
2. Gilmore  et al. 1989     ;    
3. Mera et al. 1998                ;
4. Sackett 1997              ;
5. Rana 1991                ;
6. Tammann et al. 1994      ;
7. Scalo  1986              ;
8. Kroupa et al. 1993     ;
9. Anders and Grevesse 1989 ;
10. Grevesse et al. 1996      ;
11. Cunha and Lambert 1994  ;
12. Cardelli and Federmann 1997           ;
13. Edvardsson et al 1993   ;
14. Rocha-Pinto and Maciel 1996 ;
15. Wyse and Gilmore 1995       ;
16. van der Kruit 1986      ;
17. Pagel 1997  ;
18. Kent et al. 1991 ;
19. Robin 1998
\end{table*}

\section{The evolution of the  Milky Way}

The Milky Way  is a  heterogeneous  system, with at least
three components (halo, bulge, disk) having very different
chemical, photometric and kinematical properties. A reliable model
for the evolution of the Milky Way accounting for those properties
does not exist at present. In particular, it is not clear how the
various components are related, e.g. whether the evolution of the
halo has affected in an important way the one of the bulge or the
disk (and vice versa); in fact, it seems that the halo evolution
has been completely decoupled from the one of the disk, who evolved
quite independently (e.g. Gilmore and Wyse 1998).
Neither is it clear whether there has been
significant interaction between the various parts of the disk, through
large scale gas mouvements in the radial direction. Despite the
development of various models, all these issues are still open.

For the purposes of this work we shall adopt a very simple model for the
chemical evolution of the galactic disk, considering its as an ensemble of
independently evolving, concentric, rings.
In our modelisation we are guided  by phenomenology
rather than theoretical principles and we try to construct a model
that reproduces  all the major observational
constraints of the Milky Way with a minimum number of free parameters.

\subsection{The solar neighborhood}

\subsubsection{Chemical evolution}
In the solar neighborhood (defined as a cylinder of $\sim$1 kpc radius
at a distance $R_S$=8   kpc from the galactic center), the main
observables relevant to chemical evolution are (for a detailed
discussion of these observables see  PA95 and
PS98; see also Table 1 for data and references and Fig. 4 for a graphical
presentation of the data):

i) the current surface densities of gas ($\Sigma_G$), 
stars ($\Sigma_*$) and total amount of
matter ($\Sigma_T$), as well as the current star formation
rate  $\Psi_0$; a recent analysis of all available data, 
interpreted in the framework of a consistent mass model for
the Milky Way (Mera et al. 1998), shows that there is no compelling 
evidence for the presence  of substantial amounts of sub-stellar objects
(brown dwarfs), but a limited amount ($\Sigma_{BD}<$ 8 \mp) cannot be excluded
at present.

ii) the abundances of various elements and isotopes
 at solar birth (X$_{i,\odot}$) and today (X$_{i,O}$).

iii) the age-metallicity relationship, traced by the Fe
abundance of long-lived, F-type stars.

\begin{figure*}
\psfig{file=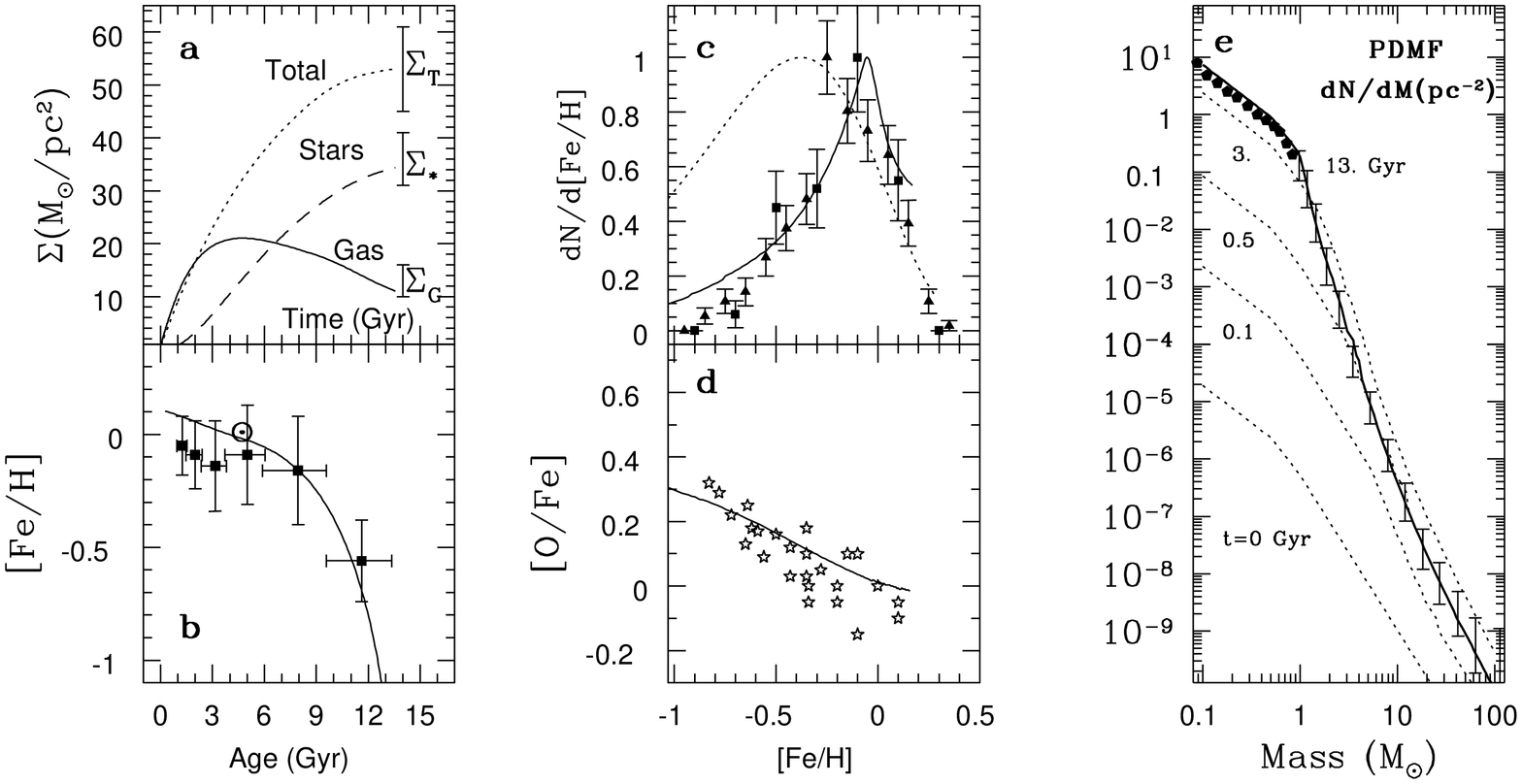,angle=-90,height=12cm,width=\textwidth}
\vspace{-3.5cm}
\caption{ \small
Results of the chemical evolution model for the solar neighborhood and comparison
to observations (see Sec. 3.1.1 and Table 1 for References).
{\bf a:} Surface densities of stars, gas and total amount of matter
as a function of time (present day data within vertical {\it error bars}; 
{\bf b:} Age-Metallicity relationship; {\bf c:} G-dwarf
differential metallicity distribution, with results from our model ({\it solid}
curve) and a closed-box model ({\it dotted } curve) shown for comparison;
{\bf d:} Oxygen vs. Fe relationship; {\bf e:} Evolution of the Present Day Mass
Function ({\it dotted } curves, at various instants) and comparison of the
final one ({\it solid} curve) to the observed PDMF of Scalo (1986, vertical
{\it error bars} for M$>$1 \ms \ stars) and of KTG93 ({\it filled symbols}, for
low mass stars);
for M$>$1 \ms \ the KTG93 IMF coincides with the one of Scalo (1986).
Obviously, at time t=0  the PDMF coincides with the adopted IMF; at t=13 Gyr
this is true for low-mass stars.
}
\end{figure*}

iv) the oxygen vs. Fe (O-Fe) relationship, interpreted in terms of a delayed
($\sim$1 Gyr) appearence of  SNIa, producing most of galactic Fe.

v) the metallicity distribution of long-lived  G-type  stars
(Fig. 4c), showing that very few of them were formed at     [Fe/H]$<$-0.7
(1/5 solar) or [O/H]$<$-0.5 (1/3 solar).

vi) the present day mass function (PDMF), resulting from the stellar IMF and the
SFR history; this observable is rarely considered in studies of the local
chemical evolution (wih the exception of Ferrini et al. 1994). However,
the resulting theoretical PDMF constitutes an important consistency check
for the adopted SFR and IMF.

The simple model decribed in Sec. 2.1 can reproduce the above constraints,
as can be seen in Figs. 4 and 5a.
The adopted SFR and infall rates
lead to a current gas surface density of $\Sigma_G\sim$10 \mp \ and
to a final stellar surface density of $\Sigma_*\sim$35
\mp, both compatible with the observations.
The final amount  of stellar remnants ($\Sigma_R\sim$7 \mp, for the sum
of white dwarfs + neutrons stars + black holes) is
slightly smaller than the current gas amount.
Models of the local chemical evolution cannot avoid producing
this amount of stellar remnants, i.e. about 15\% of the local surface density
(a different IMF would slightly change this figure).

With the adopted infall and star formation rates, a current
SFR $\sim$3 \mg \ is obtained at T=13.5 Gyr, 
again in agreement with observations.
Those ingredients, combined with the adopted IMF and stellar
yields lead to a local metallicity close to the solar one
4.5 Gyr ago and slightly higher today (Fig. 4b). It should be noted here that
observations of CNO abundances in young stars and gas in Orion show that
the metallicity of this young region is lower than solar (Cunha and Lambert 1994,
Cardelli and Federmann 1997), a result which is
difficult to interpret in conventional chemical evolution models (see e.g.
PA95); imperfect or non-instantaneous mixing could, perhaps, help explain this
observable, as well as the scatter in the age-metallicity relationship (e.g.
Coppi 1997, Thomas et al. 1998).

The current SN II rate in the solar neighborhood
is found to be compatible with the estimate of
Tammann et al. 1994 (fig. 8f): 2 10$^{-11}$ pc$^{-2}$ yr$^{-1}$.
The SNIa rate is adjusted to reproduce the observed decline of O/Fe
in the local disk (Fig. 4d). The introduction of this delayed Fe source leads to
an age-metallicity relationship somewhat steeper than 
(but still compatible with) the observations (Fig. 4b).

The differential metallicity distribution
(DMD) of G-dwarfs constitutes one of the strongest constraints for the
evolution of the solar neighborhood. It represents the number of long-lived
stars per unit logarithmic metallicity interval and can be evaluated as:
$$  \frac{dN_G}{d[Fe/H]} \ = \ \frac{\alpha_G \Psi(t) dt}{d[Fe/H]} \eqno (8)$$
where $\Psi(t)$ is the SFR at time $t$ and $dN_G=\alpha_G\Psi(t)dt$ is the
number of G-type stars born with metallicities between [Fe/H] and
[Fe/H]+d[Fe/H]; $\alpha_G$ is the fraction of G-type stars in the IMF.
   Expression (8) relates explicitly
the SFR history and the age-metallicity relationship to the DMD of G-dwarfs.
If the latter two relationships were accurately known, the local SFR could be
easily derived (Prantzos 1998a), but current observational uncertainties prevent
from   such a derivation. As can be seen from Fig. 4c
the local G-dwarf DMD is nicely reproduced
with the slow infall adopted in our model. Notice that the use of
metallicity dependent stellar lifetimes cannot solve the G-dwarf problem
(e.g. Bazan and Mathews 1990; Rocha-Pinto and Maciel 1997).

Finally, in Fig. 4e we present also the evolution of the Present Day
Mass Function (PDMF), at times 0., 0.1, 0.5, 3 and 13 Gyr from the beginning and
we compare the final result with the observed PDMF of Scalo
(1986, in the mass range   1-100 \ms) and KTG93 (low mass stars only).
In the mass range $<$1 \ms \ (where the shape of the PDMF coincides with the one of the IMF),
 the adopted IMF of KTG93 differs
from the flatter ones of  Scalo (1986), Gould et al. (1997) and Reid and Gizis (1997), 
but it is compatible with the one favoured by Haywood et al (1994);
the Salpeter IMF is steeper than any of those IMFs.
In the high mass range, the shape of the PDMF is mainly determined by the stellar
lifetimes (for small variations of the IMF slope $X$)
and it is not surprising that the results of Scalo (1986) are recovered.

\begin{figure*}
\psfig{file=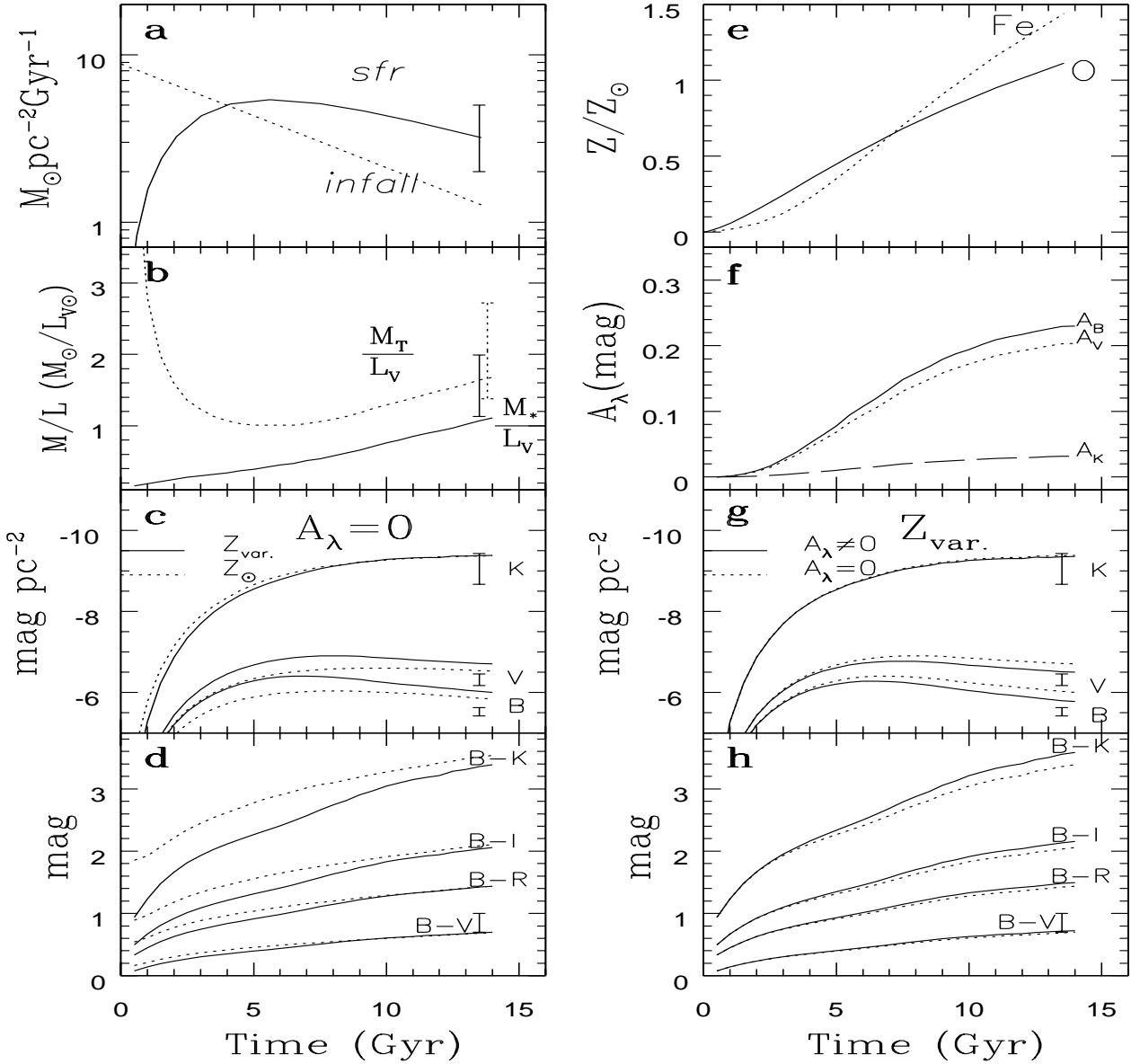,height=17cm,width=\textwidth}
\caption{ \small
Results of the chemo-photometric evolution model for the solar neighborhood (Sec. 3.1.2)
and  comparison of the values at T=13.5 Gyr to observations (with vertical {\it error bars}).
The effects of metallicity dependence of the stellar tracks are presented in pannels
{\bf c} and {\bf d}, whereas those of extinction in pannels {\bf g} and {\bf h}.
{\bf a:} Star formation and infall rates; {\bf b:} $M/L_V$ ratio for the total mass
({\it dotted} curve) and the stellar mass ({\it solid} curve); 
{\bf c:} Surface brightness in the $K, V$ and $B$ bands, obtained with the SFR of
Fig. 5a, no extinction  and stellar tracks of constant metallicity
 \zs \ ({\it dotted} curve) or variable metallicity ({\it solid} curve);
{\bf d:} Corresponding evolution of colours ({\it solid} curve: variable metallicity
stellar tracks, {\it dotted } curve: tracks with $Z=$\zs);
{\bf e:} Abundances of Fe ({\it dotted}
curve) and O ({\it solid} curve) on a linear scale, the latter been adopted as tracer
of metallicity for the calculation of extinction according to Eq. 8; {\bf f:}
Extinction in the $B, V$ and $K$ bands; 
{\bf g:} Surface brightness in the $K, V$ and $B$ bands, obtained with the SFR of
Fig. 5a, stellar tracks of variable metallicity and extinction neglected
({\it dotted} curve) or included ({\it solid} curve);
{\bf h:} Corresponding evolution of colours 
(extinction neglected: {\it dotted} curves;
 extinction included: {\it solid} curves).
{\it The solid curves in figures 5g and 5h  represent our final complete model
for the local photometric evolution}.
}
\end{figure*}

Despite the satisfactory agreement of our model with the observations,
we wish to emphasize that
the solution is by no means unique (see Tosi 1988), i.e. some other
combinations of the input parameters may also lead to acceptable
results (see Tosi 1998 for a comparison of different models).
Notice, however, that the constraints are relatively tight
and do not allow for a wild variation in the input parameters. In
particular, the metallicity distribution strongly suggests a slow
formation of the local disk; this is corroborated by the observed
current local SFR (Prantzos 1998a), which is not very different from the past
average one
(PASFR=$\Sigma_*$/T$\sim$3 \mg). A very high early SFR, declining to
very low current values, is not compatible with those data.

\subsubsection{Photometric evolution}

The set    of available observational constraints concerning the
photometric evolution of the solar neighborhood is much smaller than
the corresponding one for its chemical evolution (second part of
Table 1 and references therein). The local
surface brightness is estimated to L$_V$=22.5$\pm$3 \lv \p2.
Combined to the observed star surface density of $\Sigma_*=35 \pm 5$ \ms \p2,
it leads to a local stellar mass/light ratio of M/L$_V$=1.2-2.
(in solar units); if the total local
surface density $\Sigma_T\sim$51$\pm$5 \ms \p2 \ is adopted instead, M/L$_V$ is
found to be $\sim$1.8-2.8. Notice also the uncertainty in the local $B-V$ value, equal to
0.84$\pm$0.15 (van der Kruit 1986) or to 0.63 (Robin 1998).

The results of our model for the photometric evolution of the solar
neighborhood appear in Fig. 5. In Fig. 5b appears the evolution of the $M/L_V$ ratio
where $L_V$ is the luminosity of the stellar population  calculated with
metallicity-dependent stellar  tracks and with no extinction taken into account.
The two curves correspond to $M$ representing either the stellar mass only or
the total mass (i.e. stars + gas + stellar remnants); in both  cases, the obtained
value at T=13.5 Gyr is on the low side of the observed range.

The effects of the metallicity dependence of the adopted stellar tracks are illustrated
in Fig. 5c (luminosity evolution) and 5d (colour evolution); the results correspond
again to the luminosity of the stellar population alone, i.e. without any extinction.
It can be seen that the adoption of metallicity-dependent tracks leads to systematically
larger $B-$ and $V-$ luminosities than the case with Z=const=\zs \ tracks, especially
towards the middle of the galactic evolution where the differences can attain 0.5 mag.
Towards the end of the evolution  the differences are small, since the metallicity has increased
to $\sim$\zs  \ and the tracks with \zs \ are quite appropriate.
Notice that the final $B-$ and $V-$
luminosities are barely compatible with the observed local ones, but if extinction is
taken into account (see next paragraph) the agreement becomes satisfactory.
Notice also that the $K-$ band luminosity
is very slightly affected by these considerations. As a result, the colour evolution
of e.g. $B-K$ depends very strongly on the adopted stellar tracks. As can be seen in Fig. 5d,
the early $B-K$ is ``redder'' by almost 1 $mag$ when metallicity-independent tracks are
considered. This illustrates dramatically the importance of adopting metallicity-dependent
stellar tracks and calculating in a self-consistent way the photometric {\it and } chemical
evolution of a galaxy. On the other hand, the $B-V$ colour is very slightly affected by
the metallicity  dependence of the tracks ; the final value (0.69 with no extinction, Fig. 5d)
is quite close to the observed (but very uncertain!) value in the solar neighborhood.

\begin{table*}


{\bf Table 2:} Main observational data for the Milky Way disk and results of a simpe model
\begin{tabular}{ l l c c l }
\hline
Observable & Observed value        & References & Computed value & Main relevant parameter \\
\hline
                    &                         &  &  &   \\
{\underline {Total mass of:}} &        &  & &                            \\
 Gas                &  M$_G: $6-8 10$^9$ \ms      & 1,2,3 & 8.  10$^{9}$   & Same ingredients     \\
 Stars              &  M$_*: $4-5 10$^{10}$ \ms   &1,4    & 3.8 10$^{10}$         \\
                    &                             &       &  &                   \\
 Gas fraction       & $\sigma_G \sim$0.15         &       &   0.15   &  as for        \\
                    &                             &       &   &                  \\
 {\underline {Star formation}}                    &       &  &  &          \\
 Current SFR        & $\Psi_o$: 2-6 \my           &1,7    & 1.9  &    Solar  \\
 Frequency of SNII  & 0.55-1.0  SNU                &8,9 & 0.61  &   \\
 Frequency of SNIa  & 0.12-0.23 SNU                &8,9 & 0.16  &   \\
                    &                         &    &   &    Neighborhood   \\
 Infall rate        &   $f <$ 2 \my           &7   & 0.64  &   \\
                    &                         &    & &   \\
                    &                         &    & &         plus \\
 {\underline {Profiles}} &                    &    & &   \\
 Gas                & $\Sigma_G(R)$    &1, 2  & &   \\
 Stars              & $\Sigma_*(R) \propto$exp$^{-R/H}$&    &  &   Radially dependent   \\
                    &  ($H \sim$2.5   kpc)     &4,5,6,16  &   &                 \\
 SFR                &   $\Psi(R)$          &1, 14  &   &  SFR $\Psi(R)$ \\
                    &                      &  &   &   \\
 Abundances in      &  X$_i(R)$            &  & &         Infall $f(R)$ \\
 gas and B-stars    & d[O/H]/dR$\sim$-0.08 dex/kpc      &10,11  & -0.07  &   \\
                    &                      &  & &   \\
{\underline {Luminosities:}} &L$_B$=1.8$\pm$0.3 10$^{10}$ \lb      &12  & 1.8 10$^{10}$   & Same ingredients   \\
                             &L$_V$=2.1$\pm$1.0 10$^{10}$ \lv      &5  & 2.0 10$^{10}$  &   \\
                             &L$_K$=6.7 \ \ \ \ \ 10$^{10}$ \lk    &13  & 7.5 10$^{10}$  &  as for Solar  \\
                    &                      &  & &   \\
{\underline {Colours:}} &B-V $\sim$0.8 & 12  & 0.78  &  Neighborhood  \\
                    &                      &  & &   \\
{\underline {Scalelengths:}}&          &  & &  Photometry  \\
                    &   $H_B$=4-5 kpc      &13  & 3.9 &   \\
                    &   $H_K$=2.3-2.8 kpc  &14,15  & 2.6 &   \\
\hline
\end{tabular}

References: 
1. Prantzos and Aubert 1995; 
2. Dame 1993; 
3. Kulkarni and Heiles 1987;
4. Mera et al. 1998; 
5. Sackett 1997; 
6. Robin et al. 1992  ; 
7. Pagel 1997;
8. Tammann et al. 1994; 
9. Capellaro et al. 1997;
10. Shaver et al. 1983;
11. Smart and Rolleston 1997; 
12. van der Kruit 1986; 
13. Kent et al. 1991;
14. Wang and Silk 1994; 
15. Freudenreich 1998 ;
16. Ruphy et al. 1996

\end{table*}

On the right part of Fig. 5 are illustrated the effects of extinction on the
results of our calculation. According to the adopted prescription, 
the optical depth  depends in a sensitive way on metallicity (Eq. 5);
one should worry then about the appropriate metallicity tracer, since O and Fe
evolve in rather different ways (Fig. 5e). Since the adopted  stellar tracks
of Geneva are parametrised by the total metallicity, dominated by oxygen, we adopt
this element here as metallicity tracer in the extinction law. Another reason is that
our study concerns also other galactic regions (and, in a forthcoming paper, other spirals
as well) for which there are no data concerning their Fe evolution neither the SNIa rate
(contrary to what happens in the solar neighborhood); since the evolution of the
oxygen abundance is tidely related to the star formation rate, we feel that this element
is a safer tracer of metallicity (at least from the theoretician's point of view).

The evolution of the extinction in various wavelengths (Eq. 8) appears in Fig. 5f:
it is negligible in all wavelengths in the beginning (because of low metallicity and 
gas amount); it is always negligible in the $K-$ band, but can reach 0.3 mag in
the $B-$ and $V-$ bands.
Although the metallicity increases steadily, extinction levels out at late times, because
it depends also on the surface gas density, which decreases lately (Fig 5a).
Fig. 5g displays  the effect of extinction on luminosity: a better agreement with the
current local $V-$ and (in particular) $B-$ luminosity is obtained when extinction
is taken into account. Finally, Fig. 5h displays the corresponding evolution of
various colours; $B-V$ is not affected by extinction (since both the $B-$ and $V-$ bands
are affected in a similar way), while
$B-K$ increases by $\sim$0.25 mag.

 From Figs. 5a,g,h it becomes clear that the effect of extinction on the local photometric
evolution is small, mainly because of the relatively small current amount of gas.
The other parameters of the model (shape of the IMF, SFR history) have a stronger
impact on that evolution. Notice, however, that
this is not the case for inner galactic regions having larger gaseous and metal amounts
(see next Section).

 As a final comment, we note that the effect of metallicity dependent stellar tracks
is most important in the early galactic evolution, whereas the effects of extinction
become (relatively) important only at late times.

\subsection{The  Milky Way disk}

\subsubsection {Chemical evolution}

Contrary to the case of the solar neighborhood, the available
observations for the Milky Way disk offer information mainly
about its current status, not its past history. The main
observables relevant to chemical evolution are (see PA95
and PS98 for a more detailed discussion; also Table 2 for data and references):

\begin{figure*}
\psfig{file=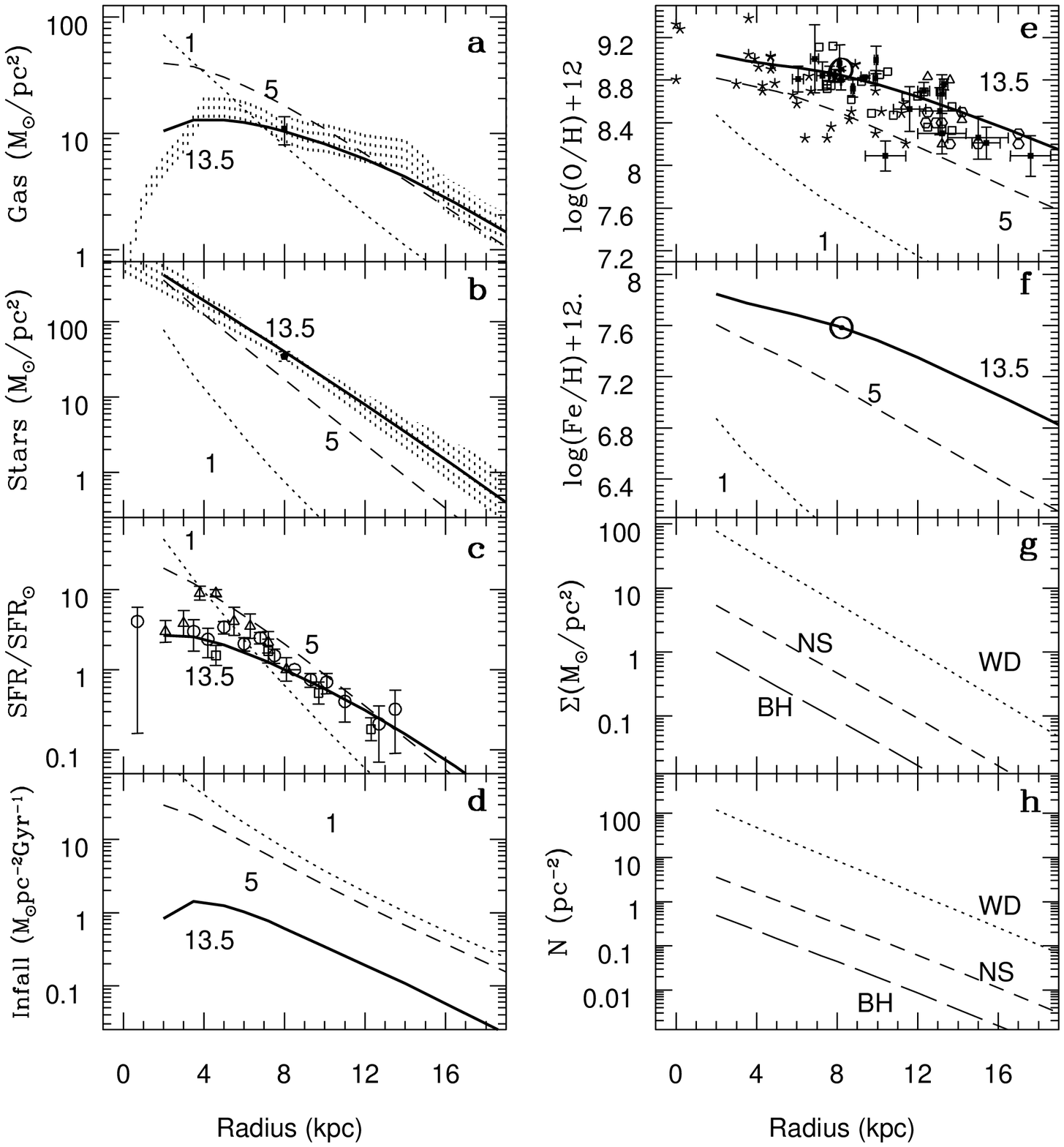,height=20cm,width=\textwidth}
\vspace*{-4cm}
\caption{ \small
Results of the chemical evolution model for the Milky Way disk (Sec. 3.2.1)
and  comparison to observations (see Table 2 for references).
{\bf a:} Gas profile at 1 Gyr  ({\it dotted} curve), 5 Gyr ({\it dashed} curve) and
13.5 Gyr ({\it thick solid} curve) and comparison of the latter to observations
of the current gaseous profile ({\it grey} area, normalised to the local
gas surface density, within {\it error bars}) ;
{\bf b:} Stellar profile 
at 1 Gyr ({\it dotted} curve), 5 Gyr ({\it dashed} curve) and
13.5 Gyr ({\it thick solid} curve) and comparison of the latter to observations
of the current stellar profile ({\it grey} area, within two exponential disks
of scalelengths 2.2 and 2.6 kpc, respectively and normalised to the current local star
surface density, within {\it error bars}) ;
{\bf c:} Star formation rate, normalised to its current value at $R_S$=8 kpc
at 1 Gyr ({\it dotted} curve), 5 Gyr ({\it dashed} curve) and
13.5 Gyr ({\it thick solid} curve) and comparison of the latter to observations 
(see Fig. 2 and references therein for observational data);
{\bf d:} Profile of the infall rate
at 1 Gyr ({\it dotted} curve), 5 Gyr ({\it dashed} curve) and
13.5 Gyr ({\it thick solid} curve) ;
{\bf e:} Oxygen abundance profile
at 1 Gyr ({\it dotted} curve), 5 Gyr ({\it dashed} curve) and
13.5 Gyr ({\it thick solid} curve) and comparison of the latter to observations
{\bf f:} Iron abundance profile
at 1 Gyr ({\it dotted} curve), 5 Gyr ({\it dashed} curve) and
13.5 Gyr ({\it thick solid} curve) ;
{\bf g:} Current surface densities (by mass) of:
white dwarfs (WD,
{\it dotted} curve),
neutron stars (NS, {\it short-dashed} curve) and black holes (BH, {\it long-dashed} curve)
{\bf h:} Current surface densities (by number) of:
white dwarfs ({\it dotted} curve),
neutron stars ({\it short-dashed} curve) and black holes ({\it long-dashed} curve).
}
\end{figure*}

i) The total mass of gas and stars in the disk (M$_G\sim$6-8 10$^9$ \ms \
and M$_*\sim$4-5 10$^{10}$ \ms, respectively); the total current SFR
($\sim$3-6 \ms \ yr$^{-1}$) and the current supernova rates
($\sim$1-2 SNII/century and $\sim$0.2-0.4 SNIa/century, respectively, as
suggested by observations of external spirals; the uncertainty on the value of the Hubble
constant and on the exact
spectral type of the Milky Way - Sb or Sbc - prevents from giving a precise value).
Notice that those quantities are often mentioned as  constraints
to (and  predictions of) one-zone models of the Galaxy, assumed to reflect
the evolution of the whole disk. This is obviously wrong, since
the disk is a heterogeneous system, as the observed gradients suggest
(see points iii-vi below).
A multi-zone model with different SFR histories in its various zones
should be obviously used. In fact, observations of external spirals
give rather the SN frequency in SNU (i.e. in number of SN per century and
per 10$^{10}$ \lb \ in Table 2); knowledge of the current L$_B$ for the Milky Way
allows then to infer its current SN frequency. However, a succesful
model of the Milky Way should reproduce in a self-consistent way {\it both }
its SN frequency and L$_B$ luminosity.

ii) The current gas profile, dominated by the molecular ring at 
galactocentric distance R$\sim$4-5 kpc and by HI of roughly constant
surface density at distances 6-14 kpc (Dame 1993 and Fig. 2).

iii) The stellar  profile, exponentially decreasing outwards.
The value of the characteristic scalelength is still under debate,
but recent studies converge towards low scale-lengths, around 2.5-3
kpc (Sackett 1997). A recent analysis of the COBE data (Freudenreich 1998)
also points to a scalelength $h$=2.6 kpc.
The combination of observables (ii) and (iii)
leads to a gas fraction profile steeply decreasing in the inner disk,
suggesting that the  star formation efficiency has been larger 
in those regions than in the outer disk.

iv)  The current SFR profile (traced by the  surface density
of pulsars and supernova remnants or the
H$_{\alpha}$ emissivity profile), strongly decreasing outwards
(Fig. 2 and references therein).
 Notice that the SFR profile does not follow 
 the molecular or the total (molecular+atomic) one, i.e. the SFR is not
 simply proportional to some power of the gaseous profile.
 
v) The current metallicity profile, usually traced by oxygen observed
in HII regions  (Shaver et al. 1983, Vilchez and Esteban 1996),
young planetary nebulae (Allen et al. 1998), and B-stars
(Smart and Rolleston 1997, Gummersbach et al. 1998),
showing a gradient of d[O/H]$\sim$-0.08 dex/kpc.

Notice that, since there are essentially no constraints on the past history of
the Milky Way disk
(i.e. no age-metallicity relations or metallicity distributions are
available for other regions)
 there is much more freedom in its modelisation
than in the case of the solar neighborhood. Still, it is meaningful
to construct models, insofar as the number of  parameters used
is considerably smaller than the constraints (i-v) above.

In our previous works (PA95, PS98) we presented a simple model of that
kind, i.e. one with the same physics as for the solar neighborhood
and with a radial dependence in the star formation rate SFR(R) and the
infall timescale $\tau$(R). As discussed in Sec. 2.1.1, the adopted radial
dependence has a physical basis (i.e. large scale instabilities in rotating
disks for the SFR and inside-out formation of the disk for the infall 
timescale $\tau$).
It turns out that with this simple parametrisation the model reproduces
reasonably well the constraints (i-v), as can be seen in Fig. 6.
Among those, the gaseous profile is the most difficult to reproduce
by models of that kind; we obtain a rather broad peak around 5 kpc, in
rough agreement with observations.
It is difficult to ask more from  such a simple model, especially since other
factors may have shaped the gaseous profile in the inner Galaxy (like e.g. the
presence of a bar inducing radial inflows and enhancing the SFR there).
Notice that this difficulty of the simple models has already been pointed
out in Wang (1990).

The model also reproduces reasonably well the total
current SFR and supernova rates (Table 2), as well as
various other quantities. This is a rather
encouraging success, since the
number of the new constraints is much larger than the number of the
new parameters. In fact, the stellar profile is essentially determined by the
boundary conditions (the adopted $\Sigma_T(R)$ profile in the normalisation
of Equ. 3), but the form of the star formation rate $\Psi(R)$ and infall timescale
$\tau$(R) (two parameters)
account then for the observed gaseous, SFR and oxygen profiles, as well as the
other results in Table 2 and Fig. 6.

\begin{figure*}
\psfig{file=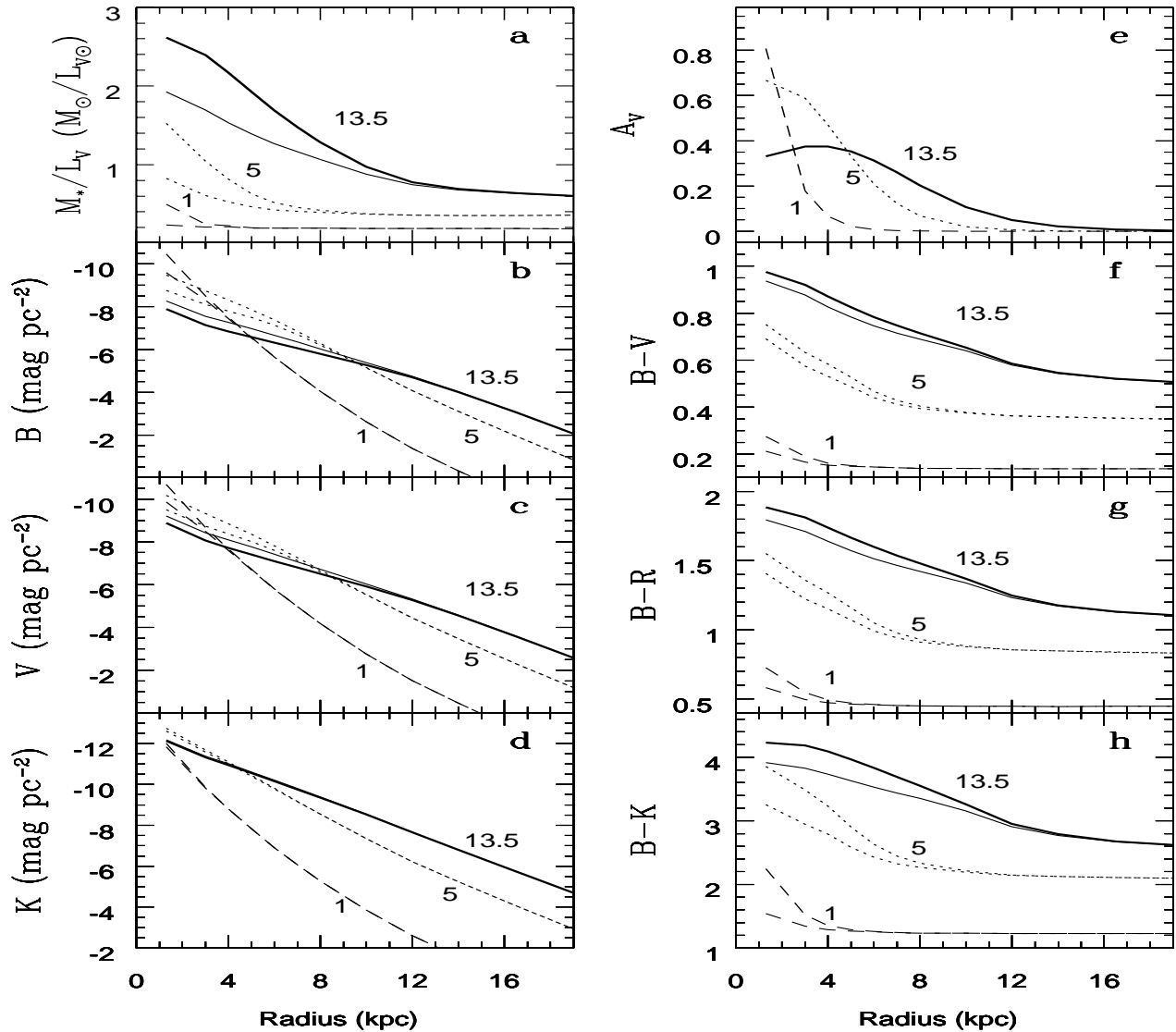,height=16cm,width=\textwidth}
\caption{ \small
Results of the chemo-photometric evolution model for the Milky Way disk (Sec. 3.2.2)
with metallicity-dependent stellar tracks.
In each panel, three curves are given, corresponding to profiles at times
t=1 Gyr ({\it dashed} curve), 5 Gyr ({\it dotted} curve) and 13 Gyr ({\it thick solid} curve),
respectively; bifurcation of the curves (mostly in the inner disk)
correspond to extinction  neglected (upper part of the bifurcation in the surface brightness
profiles, lower part in the colour profiles and $M/L_V$ ratio) or included
(lower part in the surface brightness profiles, upper part in the colour profiles  and $M/L_V$ ratio).
{\bf a:} Stellar $M/L_V$ ratio ;
{\bf b:} $B-$ surface brightness (the final exponential disk has a scalength of 
$H_B\sim$3.9 kpc
but breaks down outside $\sim$13 kpc ;
{\bf c:} $V-$ surface brightness ;
{\bf d:} $K-$ surface brightness (the final exponential disk has a scalelength
of $H_K\sim$2.6 kpc, i.e. shorter than the $B-$disk, in fair agreement with
observations ;
{\bf e:} Extinction in the $V-$ band ;
{\bf f:} Evolution of $B-V$ colour profile ;
{\bf g:} Evolution of $B-R$ colour profile ;
{\bf h:} Evolution of $B-K$ colour profile.
{\it In all the figures, the thick dotted curve presents our final complete model}.
i.e. with extinction taken into account. It is clear that extinction enhances
but does not create colour gradients, especially at late times.
}
\end{figure*}

The model can then be used with some
confidence for making further predictions. Some of them have been analysed
elsewhere (see e.g. Prantzos 1996 for the importance 
of the deuterium abundance profile
for our understanding of the past history of the disk, or Prantzos et al. 1996
on the evolution of the CO isotopic profiles). In Fig. 6 we present two more
predictions of the model concerning 
a) the evolution of the metallicity gradients in the disk
and b) the surface density of compact objects (white dwarfs,
neutron stars and black holes) as a function of
galactocentric distance.

The evolution of the metallicity gradient is an important issue
in studies of the evolution of galactic disks (see e.g. K\" oppen 1994 and
references therein). In particular, the flatenning of the metallicity
profile is supporting the idea of inside-out formation of the disk
(since the inner regions arrive to the endpoint of their evolution
more rapidly than the outer ones). In the case of our model, this
evolution is clearly seen in Fig. 6a,b,c, where the gaseous, stellar and
SFR profiles are given at three times (1, 5 and 13.5 Gyr respectively).
Unfortunately, the existing abundance data on stars and planetary nebulae
of various ages do not allow at present to  conclude on the behaviour of the
metallicity profile  on
observational basis (see Molla et al. 1997 for a thorough discussion of the
data and the associated uncertainties). The recent works of Molla
et al. (1997) and Allen et al.  (1998) conclude that the O-abundance
profile should become flatter with time and that it should
evolve very little in the past $\sim$6 Gyr, on the basis of models not very
different from ours. We confirm these conclusions, as can be seen in
Fig. 6e. Also, in agreement with Molla et al. (1997) we find that
the Fe-abundance profile is steeper than the one of oxygen at any age (Fig. 6f);
this is due to the enhanced ratio of SNIa to SNII in the inner disk, resulting
from the adopted prescription for the SNIa rate (see Sec. 2.1.4). As
already stressed in PA95, the evolution of the O vs. Fe abundance profiles
is crucial to our understanding of the past SNIa history in other regions
of the galactic disk.

The last two pannels of Fig. 6 display the final profiles of the surface
densities of stellar remnants, by mass (Fig. 6g) and by number (Fig. 6h),
respectively. As already discussed in Sec. 3.1.1, the local surface density of
stellar remnants is found to be slightly lower than the corresponding gaseous one.
Comparison to Fig. 6a shows that in the inner disk stellar remnants dominate the
gas (i.e. inside $\sim$5 kpc). In Fig. 6h it is seen that in the inner Galaxy
number densities of stellar remnants (always dominated by white dwarfs) can reach
$\sim$100 \p2. These numbers, resulting from our model, may have important implications
for a) the resulting $M/L_V$ profile (see Sec. 3.3.2); b)
statistics of experiments concerning micro-lensing events in 
the direction of the galactic bulge (e.g. Han and Chang 1998 and references therein);
or c) the detection of neutron stars
and black holes in binary systems (e.g. Romani 1998, Bethe and Brown 1998). 
The resulting total current numbers
in our model are: $\sim$10$^{10}$ white dwarfs, 5 10$^8$ neutron  
stars and$\sim$ 5 10$^7$ black holes in the Galaxy. The numbers for neutron stars
and black holes are
lower by a factor of $\sim$2  than the corresponding ones of Timmes et al. (1996),
one of the reasons being the   use of the Salpeter IMF in that work vs. the 
more realistic KTG93 IMF in our model.

\subsubsection {Photometric evolution}

As in the case of the solar neighborhood, the set of photometric data for the Milky 
Way disk is smaller (and more uncertain) than the corresponding one for chemical
evolution (Table 2). In particular, there is considerable uncertainty about the
scale-length in the various wavelengths; it is clear, however, that scalelengths
are larger in the shorter wavelengths (e.g. $\sim$4-5 kpc in the $B-$ band vs.
$\sim$2.3-3 kpc in the $K-$ band). One of the main results of this section
concerns precisely this issue.

The evolution of the luminosity profile in several wavelengths and of the associated
colour profiles appears in Fig. 7, for three different epochs: 1 Gyr, 5 Gyr and 13.5 Gyr,
 so that a direct  comparison is possible with the corresponding profiles of gas, stars,
 SFR and metallicity (Fig. 6). Our calculations are performed with metallicity dependent
stellar tracks and results are displayed with and without extinction.

The luminosity profiles clearly reflect the inside-out formation of the disk,
one of the basic ingredients of our model: in all wavelengths the disk is smaller
and more compact in early times. The $K-$ band profile reflects better the stellar
profile of Fig. 6b than the shorter wavelengths; its scalelength increases to a final
value of $\sim$2.6 kpc, in fair agreement with the observationally deduced one
(Table 2). On the other hand, the $B-$ profile is generally flatter, in particular
in the inner disk and follows closely (albeit not perfectly) the SFR profile (Fig. 6c);
both profiles reflect the young stelar population and flatten in the inner disk
at late times because of the exhaustion of gas supply in that region.
The final profile of $L_B$ is exponential, with a scalelength $\sim$4 kpc between
3 and 13 kpc, i.e. for about 3 scalelengths. Outside 13 kpc the final
$B-$ profile steepens, as does the corresponding SFR profile (Fig. 6c).

The stellar $M/L_V$ profile (Fig. 7a) has a uniform value $\sim$0.2 (in solar units)
at t=1 Gyr; its value
rises more rapidly in the inner disk (where it reaches a value of $\sim$2 at 13.5 Gyr), than
in the outer disk. As with extinction and colour profiles (see next paragraphs)
the final $M/L_V$ profile is very flat outside $\sim$13 kpc. The reason is that
in the inner disk, a large number of the stars emitting in the $V-$ band are
created early on and are dead by t=13.5 Gyr, contributing to $M$ but not to $L_V$.
In the outer disk, most of those stars
are created relatively late (because of the inside-out star formation)
and are still shining.

Extinction (Fig. 7e) plays some role in the inner disk, where $A_V$ can reach $\sim$0.4 mag
(compared to $\sim$0.2  mag in the solar neighborhood; see Fig. 5f).
Contrary to the solar neighborhood, where $A_V$ remains $\sim$constant in the past
$\sim$5 Gyr (since the effect of gas depletion is compensated by a small increase in
metallicity), in the inner disk $A_V$ {\it decreases} with time, since
the important gas depletion is not compensated by a corresponding metallicity increase
(metallicity saturates in the inner  disk lately, as seen in Fig. 6e).

In Figs. 7fgh it can be seen that colour gradients are established early on in the
inner disk and propagate outwards. Extinction modifies slightly these profiles
(especially the $B-K$ one), more at intermediate times than in the end of the
evolution (for reasons explained in the previous paragraph).
However, we find that extinction is not the main factor in shaping those
gradients, in agreement with the analysis of Kuchinski et al. (1998) for
the disks of spiral galaxies.
 The final colour profile presents an important gradient, extending up to 12-13 kpc.
 This is due to the fact that, inside that radius there are important differences
in the star formation time-scales between e.g. 2-3 kpc and  6-7 kpc; inner regions
have considerably older stellar populations than outer ones. However, outside the
12 kpc radius the disk is still young (star formation happens lately because of
the long infall timescales) and colour gradients are not established.

\begin{figure*}
\psfig{file=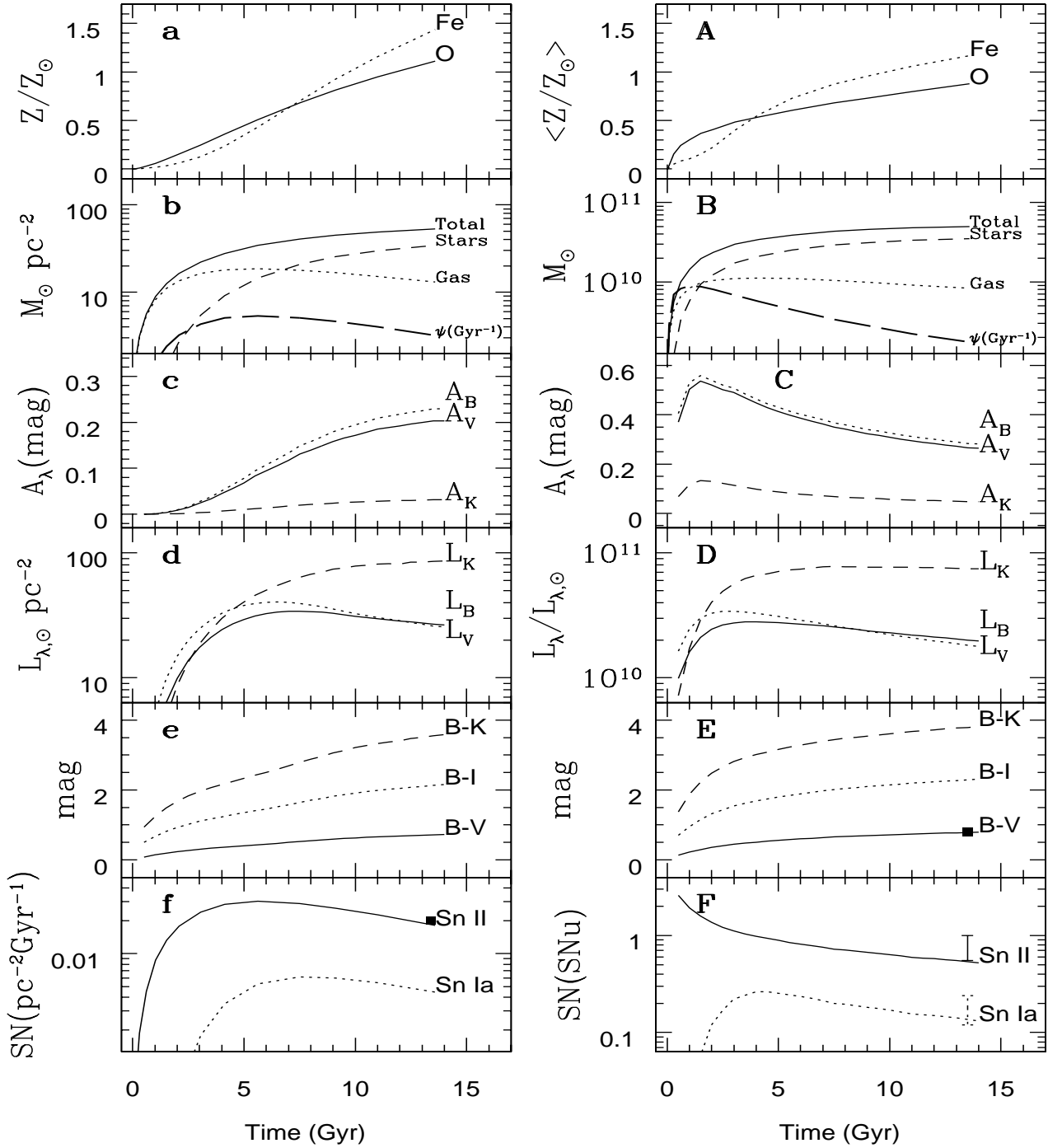,height=20cm,width=\textwidth}
\caption{ \small
Comparison of the results for the chemo-photometric evolution
of the solar neighborhood (on the left, see Sec. 3.1) and the one of the Milky Way
(on the right, see Sec. 3.3 for the calculation of integrated or average properties).
{\bf a} and {\bf A}:  Evolution of Fe and O abundances on a linear scale;
{\bf b} and {\bf B}:  Evolution of gaseous, stellar and total mass (in \mg \ on the left, in
\ms on the right), as well as of star formation rate (corresponding units per Gyr);
{\bf c} and {\bf C}:  Evolution of extinction in the $B-, V-$ and $K-$ band ;
{\bf d} and {\bf D}:  Evolution of $B-({\it dotted}), V-$ ({\it solid})
 and $K-$ ({\it dashed}) luminosities, with extinction included;
{\bf e} and {\bf E}:  Evolution of $B-K$, $B-I$ and $B-V$ colours (with extinction);
{\bf f} and {\bf F}:  Evolution of supernova rates; the local ones (Fig. {\bf 8f})
are expressed in units of pc$^{-2}$ Gyr$^{-1}$, while the total ones (Fig. {\bf 8F})
in SNu, i.e. the total SN rates (per century) 
are divided by the $B-$ band luminosity, expressed in 10$^{10}$ $L_{B_{\odot}}$. For
observational data ({\it filled squares} or {\it vertical error bars} at T=13.5 Gyr)
see Table 2 and references therein.
}
\end{figure*}

The main point of this section concerns precisely the prediction of colour
gradients in the inner disk, related to the existence of different scalelengths
for the various wavelength bands. These features are a direct consequence of
the inside-out formation scheme adopted here, on the basis of some theoretical
arguments. A ``uniform'' formation of the disk (i.e. with the same timescale
of star formation everywhere) would lead to similar scalelengths in all wavelengths
and would create very small
colour and metallicity
gradients. On the other hand, extinction may certainly enhance colour gradients, but
to a relatively small degree  (at least with the prescriptions adopted in this work).
The available data for the Milky
Way  allows only a very limited check of these ideas, but the large sample
of observations concerning external spiral galaxies offers the opportunity
for a detailed comparison; these issues are explored in a forthcoming paper
(Boissier and Prantzos 1998).

\subsection {Milky Way vs. solar neighborhood }

In studies of the photometric (and, sometimes, chemical) 
evolution of spiral galaxies with one zone models, calibration is often made
to the solar neighborhood observables, i.e. it is assumed that the local
disk evolution is representative of the average Milky Way disk evolution.
In this section we check this assumption, in the framework of our model.
We calculate total (extensive) quantities  $Q_T$ for the disk as:
$$Q_T(t) \ = \ \int_0^{R_G} \ 2 \ \pi \ q(R,t) \ R \ dR \eqno(9) $$
where $q(R,t)$ is any quantity expressed
in units of surface density (\p2) and $R_G$ is the outer galactic radius.
With this equation we obtain
$L_{\lambda}^{0}$ when integrating the stellar luminosity alone, and $L_{\lambda}$
when integrating the  luminosity after correction for extinction at each radius.
The ``integrated'' extinction is then $A_{\lambda}$=-2.5 log($L_{\lambda}$/$L_{\lambda}^{0}$).
Magnitudes and colours of the whole disk are computed from the integrated spectrum.
We also calculate average (intensive) quantities
as e.g. abundances. In this case, the average galactic value $<X_i>$ is
$$  <X_i(t)>\ = \ \frac{ \int_0^{R_G} \ 2 \ \pi \ X_i(R,t) \ \Sigma_{gas}(R,t) \  R \  dR}
{\int_0^{R_G} \ 2 \ \pi \  \Sigma_{gas}(R,t) \ R \ dR } \eqno (10) $$

The results are plotted in Fig. 8, where the solar neighborhood evolution
(left part) is compared to the one of the Milky Way disk (right part).
The final average Fe and O abundances in the gas of the disk (Fig. 8A)
are similar to the ones in the solar neighborhood (Fig. 8a), 
but their history is different:
early on, the average disk abundances are dominated by the inner galaxy which evolves
more rapidly than the solar neighborhood; thus, the average disk metallicity increases
more rapidly than the one of the local disk early on, whereas at late times
the trend is inversed.
The same behaviour is shown by the mass in stars and the SFR (Fig. 8b and 8B).

\begin{figure}
\psfig{file=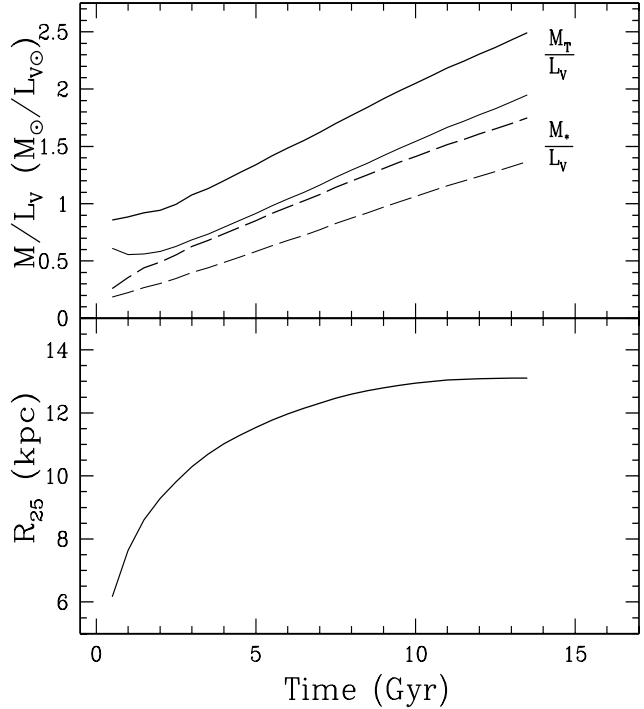,height=10.cm,width=0.5\textwidth}
\caption{\small {\it Upper pannel:} 
Evolution of the galactic M/L$_V$ ratio for the total
mass ({\it solid} curves) and the stellar mass ({\it long dashed } curves), with
extinction neglected (i.e. stellar population alone, {\it thin} curves) or
included ({\it thick} curves).
{\it Lower pannel:} 
Evolution of the isophotal radius R$_{25}$,
at which the $B$-band surface brightness is equal to 25 mag arcsec$^{-2}$. }
\end{figure}

An issue of considerable interest is the one of the SFR vs. gas amount.
In the case of the solar neighborhood, the current/maximum ($q_{now}/q_{max}$)
ratio is $\sim$0.7
for the gas surface density and $\sim$0.5 for the SFR density (as expected from
the $\Psi \propto \Sigma_G^{1.5}$ law adopted locally); both the gas and the SFR have
a broad maximum around 5-6 Gyr. In the case of the Milky Way, the corresponding
ratio is $\sim$0.8 for the total gas and $\sim$0.20 for the total SFR, i.e.
those two global quantities do not obey the  local Schmidt law; besides, their
maxima do not coincide, being at $\sim$3-4 Gyr for the gas and $\sim$1-2 Gyr for
the SFR. The reason of this behaviour is, of course, the non-linearity of the
adopted SFR law, due both to the exponent (1.5) and the $R^{-1}$ factor; both
these factors enhance the SFR efficiency early on in the inner disk, where there
is a lot of gas infalling rapidly. At late times, even if the total amount of gas
is the same, it is distributed differently. A large part of the total SFR comes now
from the outer regions (favoured by their larger aera), where the SFR efficiency
is small. In view of the importance of that topic, we make a more detailed analysis
in Sec. 3.4, where a comparison to observational data of external spirals is performed.

In Fig. 8c and 8C it can be seen that are also important differences between the
Solar neighborhhod and the Milky Way concerning the evolution of extinction.
In the former case, $A_B$ and $A_V$ rise steadily, while in the latter they undergo
an early maximum (due to the combined large amounts of gas {\it and} metals in the
inner disk) and then decline steadily, since the gas in the inner regions is
consummed and its metallicity barely increases; the outer regions play no role
in the overall extinction, in view of their low column densities and metallicities.

The integrated luminosity of the Galaxy (Fig. 8D) evolves strongly in the first several
Gyr, being dominated by the inner disk. After that period,
the $K$-luminosity stays nearly constant (most of the galactic stars are already formed),
while the $ B-$ and $V-$ luminosities decrease
slowly (because of the declining total SFR of Fig. 8B, although they do not follow it
perfectly). At $R_S$=8 kpc (Fig 8d), luminosities rise and decline more slowly.
This difference in time-scales is also present in the colour evolution (fig. 8e and 8E):
while the solar neighborhood  reddens steadily from t=0 to the present time, the integrated
colours have an evolution which is more rapid at early times and slower at late times.

The SN rates for the whole disk (fig. 8F) are expressed
in SNU (SN per century per 10$^{10}$ \lb).
The SNII frequency (in SNu) is high early on and decreases steadily, because
$L_B$ does not follow exactly the SFR (stars contributing to $L_B$ are not as short lived
as those exploding as SNII).
The results at T=13.5 Gyr for both SNII and SNIa
are closer to the rates  reported (Tammann et al. 1994, Capellaro et al. 1997)
for Sb rather than Sbc galaxies and for low rather than high values of the Hubble constant.
(both factors, i.e. the Sb type and low Hubble value,  lead to lower deduced frequencies in SNu).

Finally, the evolution of the $M/L_V$ ratios, obtained by dividing the stellar and
total mass of the disk (fig. 8B) by the $L_V$ luminosity (fig. 8D), is shown in figure 9.
Both quantities increase steadily and more rapidly than in the case of
the solar neighborhood (Fig. 5b), because most of the disk mass resides in the inner
galaxy, which evolves more rapidly than the local disk. Also, the M/L$_V$ ratio
depends somewhat on the calculated amount of
extinction (up to 45\% in the early evolution and to 20\% towards the end).

In summary, the overall evolution of the Milky Way disk bears little similarity to
that of the solar neighborhood (or to any other region for that matter). Because
of the non-linearity of the SFR efficiency with galactic radius, evolution
is dominated by the inner disk at early times. At late times,  most of the ``action''
takes place in the outer disk, which dominates extensive quantities
(Equ. 9) that follow the SFR; at those late times,
intensive quantities (Equ. 10) receive important contributions from the whole disk.
This behaviour of spiral disks resulting from our model
has important implications  for the observations of external spirals, both at
low and high redshift.

\begin{figure}
\psfig{file=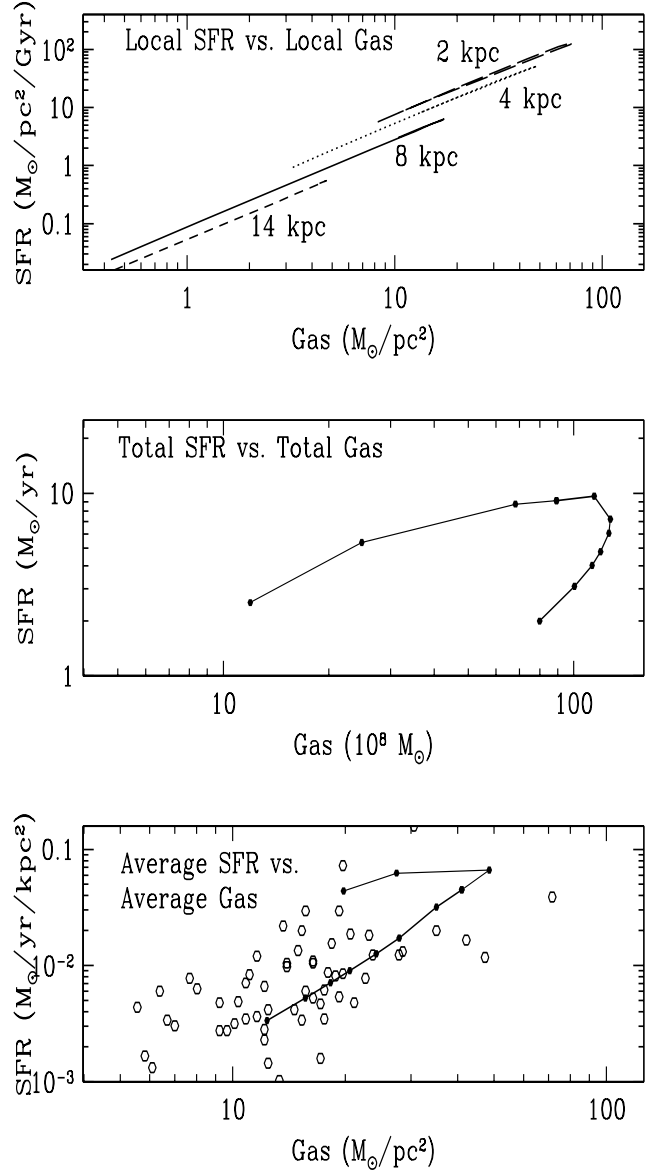,height=17.cm,width=0.5\textwidth}
\caption{ \small
{\it Upper pannel:} Local relationship between SFR and gas surface densities
for four zones of our model ({\it curves}) during the galactic evolution; 
for each zone, a unique  Schmidt law with slope 1.5
is recovered, but the shift between the curves indicates that the $R^{-1}$ factor
increases the SFR efficiency in the inner regions.
{\it Middle pannel:} Relationship between the total SFR and gas amount of the disk;
evolution takes place from left to right  in this diagram and points mark various
instants in time (0.01, 0.1, 0.5, 1., 2., 4., 5.5, 7.5, 9., 11. and 13.5 Gyr, respectively).
The local Schmidt law is not recoverd globally, because of
the non-linearity of the SFR across the disk (see the discussion in Sec. 3.3).
{\it Lower pannel:} Relationship between the average SFR and average 
gas surface density in the disk;
the quantities of the previous pannell  are divided by $\pi$R$_{25}^2$, where
R$_{25}$ is the
radius at which $\mu_B$ = 25 mag arcsec$^{-2}$. Results
({\it solid} curve with points marking several instants in time) are compared
to observations in external spirals ({\it open} symbols); the data are from Kennicutt
(1998), but the given amount of hydrogen (H$_2$+HI) has been increased  by 40\%, to take He
into account, as to be directly comparable with our results.}
\end{figure}

\subsection  {The average SFR in the Milky Way and normal spirals}

The  behaviour of local vs. global  SFR, briefly discussed in the previous Section,
is better illustrated
in Fig. 10.  The local SFR in each galactic zone obeys the adopted Schmidt law
(upper pannel), but with considerably
different efficiencies (because of the $R^{-1}$ factor).
However, the global SFR vs. total gas amount (middle pannel in Fig. 10) has a completely
different behaviour: for the same gas amount, very different values of the SFR
are found. It is true that the high values concern the early evolution (before the
first 2 Gyr), where the SFR is dominated by the innermost regions of the disk
and the adopted SFR prescription is probably not valid (since there are no spiral
arms inside $\sim$2 kpc, at least in the case of the current Milky Way).
When the first 2 Gyr are neglected, we find a relationship between global SFR and
gas amount (thick part of the curve in Fig. 10, middle), which corresponds to a slope
of $\sim$4, i.e. SFR$_{TOTAL} \propto$ M$_{GAS}^4$; this simply reflects the highly
non-linear behaviour of the adopted SFR prescription, but {\it has no physical meaning} 
and cannot be used in any model of chemical evolution. Indeed,
the important conclusion of this analysis is that, even if a Schmidt type SFR law
is  introduced locally, any radial dependence of the SFR efficiency  will make
the global SFR of a disk deviate considerably from this law. In other terms,
{\it it is impossible to model spiral galaxies by one zone models with a fixed
relationship between SFR and gas amount}. This would have some sense only if there
were no radial variation in the SFR efficiency, but in that case {\it no gradients
of abundances or colours would be obtained}, contrary to observational evidence.

A more physical insight is obtained by analysing the behaviour of {\it average}
quantities, i.e. average SFR density vs. average  gas density ; both averages are made
over some portion of the disk area. In a recent study 
concerning  normal spiral galaxies and starbursts
Kennicut (1998) divides
total SFR and $\Sigma_{GAS}$ by
$\pi \ R_{25}(B)^2$, where $R_{25}(B)$ is the
radius at which the surface brightness in the $B$-band is equal to 25 mag arcsec$^{-2}$.
The evolution of $R_{25}(B)$ in our model is given in Fig. 9 (lower pannel):
it increases in the first 6
Gyr up to R$_{25}$ $\sim$ 13.5 kpc (because of the adopted infall and SFR laws)
and stays roughly constant afterwards.  By dividing the model total SFR and $\Sigma_{GAS}$
by $\pi \ R_{25}(B)^2$, we are able to
compare directly our results to the data of
Kennicutt (1998). His data for
normal spirals are plotted also in Fig. 10 (lower pannel), but we increased his
gas surface densities (concerning the sum of H$_2$+HI) by 40\% (=1./0.7)
to account for
a contribution of 30\% He by mass (to compare directly with our results).
Our model curve (after the first 0.5 Gyr) lies well within the data points, i.e.
the {\it absolute} values of the average SFR  during the Milky Way evolution
correspond fairly well to observations of external spirals. Also, after the first 0.5 Gyr
we find a unique slope N=1.7
in the average SFR vs. gas surface density relationship. This value is to be
compared with the values derived by Kennicutt (1998): N=1.29$\pm$0.18 for a conventional
least squares fit to his data
and N=2.47$\pm$0.39 for a bivariate least squares regression,
taking into account uncertainties in both the SFR and the gas density.
Kennicutt (1998) concludes that ``...any Schmidt type law in these galaxies
should be regarded as a very approximate parametrisation at best''. We showed here
that the adopted local Schmidt law leads to an average SFR compatible with
available data for external spirals. However, this does not imply that
the adopted N=1.5 is the real value of the local Schmidt law; our result only
implies that, when combined with the $R^{-1}$ factor for the SFR efficiency,
the N=1.5 exponent leads
to results that are compatible with all available observables in the Milky Way
and with observations of the average SFR vs. gas density in
other spirals. Some other combination (i.e. a different N and a different
radial dependence of the SFR efficiency) could, perhaps, lead also to acceptable
results.

\section {Conclusions}
%

We have developed a model computing coherently the chemical
\emph{and} spectrophotometric evolution of spiral disks. 
The model makes use of up-to-date input ingredients (i.e. stellar IMF and yields,
metallicity dependent stellar lifetimes, evolutionary tracks and spectra)
and considers the galactic disk as an ensemble of independently evolving concentric rings
built by infall of gas of primordial composition.
Its main ingredient is a radially dependent SFR ($\Psi(R) \propto \Sigma_G^{1.5} R^{-1}$),
based both on empirical data (the $\Sigma_G^{1.5}$ part) and theoretical arguments (the
$R^{-1}$ part). In fact, we show here (Sec. 2.1.3 and Fig. 2) 
that the adopted SFR law, when applied
to the gaseous profile of the Milky Way, gives results which compare fairly well with the
observed SFR profile in our Galaxy.
The model is then applied to the Milky Way evolution and
the main results can be summarised as follows:

1) The main observational features of the Solar neighborhood and the Milky Way disk 
are fairly well
reproduced, with the simple assumptions of   a slow formation of the local disk (in timescales
of many Gyr, in order to explain the observed local G-dwarf metallicity distribution) 
and a radial variation in the efficiency of star formation 
(to obtain the observed gas and abundance gradients).

2) The evolution of the abundance gradients is not really constrained by observations at present.
In our  model, abundance gradients are predicted to flatten with time and to finally saturate 
in the inner disk (in agreement with a few previous works on that topic). Also, the adopted
prescription for the rate of SNIa (major Fe producers) reproduces successfully  a local
observable (the decline of O/Fe with Fe/H in the solar neighborhood) and predicts
a gradient of Fe/H steeper than the one of oxygen at all times (Sec. 3.2.1).

3) The predicted current SNII rate in the solar neighborhood is in fair agreement with
observations (2 10$^{-11}$ core collapse SN pc$^{-2}$ yr$^{-1}$, Tamman et al. (1994)),
showing that our choice of SFR and IMF are mutually consistent; this is also
coroborated by the fact that the model reproduces fairly well the local
Present Day Mass Function, something rarely considered in analogous studies (Sec. 3.1.1).
More importantly, we obtain the total current rate of SNII and SNIa expressed in SNU, i.e.
by calculating both the supernovae rates and the evolution of the blue luminosity of the
Galaxy. The results are on the low range of (but compatible with) observations of
external galaxies (Sec. 3.3 and Table 2).

4) We calculate the surface density profiles of the various stellar remnants (dominated
always by white dwarfs). We find that currently in the solar neighborhood their contribution is
smaller than the one of gas, while in the inner disk it is much larger and can reach $\sim$20\%
of the stellar surface density (Sec. 3.2.1).
We find that there should currently exist $\sim$10$^{10}$ white dwarfs, $\sim$5 10$^8$ neutron
stars and $\sim$5 10$^7$ black holes in the Galaxy (within a factor of $\sim$2, depending
on the IMF and the mass of the disk).
On the total, the galactic mass of compact objects is comparable to the one of gas, i.e.
each contributes for $\sim$7-8 10$^9$ \ms \ or $\sim$12-15\% of the mass of the disk.
These results may have important implications for the statistics
of microlensing events towards the bulge, or for the detection of galactic
neutron stars and black holes in binary systems.

5) The use of metallicity dependent stellar tracks, lifetimes and spectra, along with a full
chemical evolution model, is mandatory in studies of the photometric evolution of galaxies
(especially of systems with large metallicity variations in long timescales, such as
 spirals, see Sec. 3.1.2). We also show that, when metallicity dependent lifetimes are used
(as they should!) only the isochrone method is applicable when calculating the
luminosity evolution of a galaxy (Sec. 2. 2).

6) The adopted scheme of star formation in the disk leads naturally to different scalelengths
in the various photometric bands (shorter in the red and longer in the blue), in fair
agreement with observations of the Milky Way disk for which we obtain: $H\sim$2.6 kpc in the $K$ band
and $H\sim$4 kpc in the $B$ band (Sec. 3.2.2).
Also, it produces galactic disks which are much more compact in the past,
an important result in view
of current studies of galaxy evolution at high redshifts.

7) For the same reason, colour gradients are obtained, first in the inner disk, then propagating
outwards. We find no colour gradients in the outer disk,
which is formed relatively late in that scheme, so that there is no time for 
colour gradients to be established there (Sec. 3.3.2).
Moreover, we find that extinction enhances but does not really create colour gradients
(at least with the adopted prescription).

8) The evolution of various parameters in the solar neighborhood  
(stars, gas, abundances, luminosities etc.)
does not match the corresponding evolution (of average or global  quantities) of the whole disk.
This is due to the non-linearity of the adopted star formation law 
as a function of galactocentric radius.
In particular, the early evolution of the Galaxy (dominated by the inner disk)
is more rapid and the late evolution slower than
the one of the solar neighborhood. This implies that one-zone models reproducing the evolution
of the solar neighborhood cannot be taken as representative of the evolution of the Milky Way
as a whole (Sec. 3.3).

9) The adopted scheme of star formation (main ingredient of the model) compares fairly well
to observations of average SFR vs. gas surface density of external spirals. Notice that such
a comparison can be made only in the framework of  chemo-photometric models 
since the R$_{25}$ radius is required for the averaging (see Sec. 3.4)

We notice that the photometric results of our model depend strongly on the quality of the
adopted stellar tracks and spectra. In that respect, the use of the now available 
 homogeneous set of
these ingredients presents a great improvement. However, the Geneva tracks include neither
the Horizontal Branch (negligible in the context of this work) nor the thermally pulsing
AGB phase (potentially more important, but certainly not able to invalidate our conclusions
in points 6 and 7). 

Finally,
the success of the model does not necessarily imply its correctness. It suggests however
that there may be a grain of truth in the overall picture, which should be further tested
against a larger mass of observational data, concerning spiral galaxies at low and high redshifts.
Work is in progress along these directions.

\label{lastpage}

\end{document}